\input lanlmac
\input amssym

\yearltd=\year

\font\tenmsb=msbm10
\font\ninemsb=msbm9
\font\sevenmsb=msbm7
\font\fivemsb=msbm5
\def\bb{
\textfont1=\tenmsb
\scriptfont1=\sevenmsb
\scriptscriptfont1=\fivemsb
}
\let\oldfoot\foot
\long\def\foot#1{\oldfoot{\def\bb{%
\textfont1=\ninemsb
\scriptfont1=\sevenmsb
\scriptscriptfont1=\fivemsb
}#1}}

\def\I{\hbox{$\bb I$}}

\def\R{\hbox{$\bb R$}}
\def\C{\hbox{$\bb C$}}
\def\H{\hbox{$\bb H$}}
\def\Z{\hbox{$\bb Z$}}
\def\A{\hbox{$\bb A$}}
\def\S{\hbox{$\bb S$}}
\def\L{{\cal L}}
\def\M{{\cal M}}
\def\N{{\cal N}}
\def\T{{\cal T}}
\def\U{{\cal U}}
\def\tr{\hbox{tr}\,}
\def\grav{{\rm grav}}
\def\vev#1{\langle#1\rangle}
\def\emph#1{{\it #1}}
\def\Nequals#1{$\N{=}#1$}

\def\Arf{{\rm Arf}}
\def\Hom{{\rm Hom}}

\def\G{{\cal G}}
\def\m{{\bf m}}

\lref\WittenNI{
  E.~Witten,
  ``The Central Charge In Three-dimensions,''
  in Physics and Mathematics of Strings: Memorial Volume for Vadim Knizhnik, pp.~530--559, World Scientific, 1990.
}

\lref\WittenGF{
  E.~Witten,
  ``On S duality in Abelian gauge theory,''
Selecta Math.\  {\bf 1}, 383 (1995).
\href{https://arxiv.org/abs/hep-th/9505186}{[hep-th/9505186]}.
}

\lref\DineXK{
  M.~Dine, N.~Seiberg and E.~Witten,
  ``Fayet-Iliopoulos Terms in String Theory,''
Nucl.\ Phys.\ B {\bf 289}, 589 (1987)..
}

\lref\KomargodskiPC{
  Z.~Komargodski and N.~Seiberg,
  ``Comments on the Fayet-Iliopoulos Term in Field Theory and Supergravity,''
JHEP {\bf 0906}, 007 (2009).
\href{https://arxiv.org/abs/0904.1159}{[arXiv:0904.1159 [hep-th]]}.
}

\lref\KomargodskiRB{
  Z.~Komargodski and N.~Seiberg,
  ``Comments on Supercurrent Multiplets, Supersymmetric Field Theories and Supergravity,''
JHEP {\bf 1007}, 017 (2010).
\href{https://arxiv.org/abs/1002.2228}{[arXiv:1002.2228 [hep-th]]}.
}

\lref\SeibergQD{
  N.~Seiberg,
  ``Modifying the Sum Over Topological Sectors and Constraints on Supergravity,''
JHEP {\bf 1007}, 070 (2010).
\href{https://arxiv.org/abs/1005.0002}{[arXiv:1005.0002 [hep-th]]}.
}

\lref\WittenYA{
  E.~Witten,
  ``SL(2,Z) action on three-dimensional conformal field theories with Abelian symmetry,''
in From fields to strings, Memorial Volume for Ian Kogan, vol.~2, pp.~1173--1200, World Scientific, 2005.
\href{https://arxiv.org/abs/hep-th/0307041}{[hep-th/0307041]}.
}

\lref\KapustinGUA{
  A.~Kapustin and N.~Seiberg,
  ``Coupling a QFT to a TQFT and Duality,''
JHEP {\bf 1404}, 001 (2014).
\href{https://arxiv.org/abs/1401.0740}{[arXiv:1401.0740 [hep-th]]}.
}
\lref\GaiottoKFA{
  D.~Gaiotto, A.~Kapustin, N.~Seiberg and B.~Willett,
  ``Generalized Global Symmetries,''
JHEP {\bf 1502}, 172 (2015).
\href{https://arxiv.org/abs/1412.5148}{[arXiv:1412.5148 [hep-th]]}.
}
\lref\GaiottoYUP{
  D.~Gaiotto, A.~Kapustin, Z.~Komargodski and N.~Seiberg,
  ``Theta, Time Reversal, and Temperature,''
JHEP {\bf 1705}, 091 (2017).
\href{https://arxiv.org/abs/1703.00501}{[arXiv:1703.00501 [hep-th]]}.
}

\lref\TachikawaAUX{
  Y.~Tachikawa and K.~Yonekura,
  ``Anomalies involving the space of couplings and the Zamolodchikov metric,''
JHEP {\bf 1712}, 140 (2017).
\href{https://arxiv.org/abs/1710.03934}{[arXiv:1710.03934 [hep-th]]}.
}

\lref\GaiottoYUP{
  D.~Gaiotto, A.~Kapustin, Z.~Komargodski and N.~Seiberg,
  ``Theta, Time Reversal, and Temperature,''
JHEP {\bf 1705}, 091 (2017).
\href{https://arxiv.org/abs/1703.00501}{[arXiv:1703.00501 [hep-th]]}.
}

\lref\Mumford{
D.~Mumford,
``Stability of projective varieties,''
L'Enseignement Math\'ematique {\bf 23}, (1977) 39--110.
}

\lref\GomisYAA{
  J.~Gomis, P.~S.~Hsin, Z.~Komargodski, A.~Schwimmer, N.~Seiberg and S.~Theisen,
  ``Anomalies, Conformal Manifolds, and Spheres,''
JHEP {\bf 1603}, 022 (2016).
\href{https://arxiv.org/abs/1509.08511}{[arXiv:1509.08511 [hep-th]]}.
}

\lref\SeibergRSG{
  N.~Seiberg and E.~Witten,
  ``Gapped Boundary Phases of Topological Insulators via Weak Coupling,''
PTEP {\bf 2016},  12C101 (2016).
\href{https://arxiv.org/abs/1602.04251}{[arXiv:1602.04251 [cond-mat.str-el]]}.
}

\lref\DonagiVWH{
  R.~Donagi and D.~R.~Morrison,
  ``Conformal field theories and compact curves in moduli spaces,''
\href{https://arxiv.org/abs/1709.05355}{[arXiv:1709.05355 [hep-th]]}.
}

\lref\SeibergGMD{
  N.~Seiberg, T.~Senthil, C.~Wang and E.~Witten,
  ``A Duality Web in 2+1 Dimensions and Condensed Matter Physics,''
Annals Phys.\  {\bf 374}, 395 (2016).
\href{https://arxiv.org/abs/1606.01989}{[arXiv:1606.01989 [hep-th]]}.
}

\lref\MetlitskiYQA{
  M.~A.~Metlitski,
  ``$S$-duality of $u(1)$ gauge theory with $\theta =\pi$ on non-orientable manifolds: Applications to topological insulators and superconductors,''
\href{https://arxiv.org/abs/1510.05663}{[arXiv:1510.05663 [hep-th]]}.
}

\lref\GaiottoWE{
  D.~Gaiotto,
 ``N=2 dualities,''
JHEP {\bf 1208}, 034 (2012).
\href{https://arxiv.org/abs/0904.2715}{[arXiv:0904.2715 [hep-th]]}.
}

\lref\AharonyHDA{
  O.~Aharony, N.~Seiberg and Y.~Tachikawa,
  ``Reading between the lines of four-dimensional gauge theories,''
JHEP {\bf 1308}, 115 (2013).
\href{https://arxiv.org/abs/1305.0318}{[arXiv:1305.0318 [hep-th]]}.
}

\lref\TachikawaHYA{
  Y.~Tachikawa,
  ``On the 6d origin of discrete additional data of 4d gauge theories,''
JHEP {\bf 1405}, 020 (2014).
\href{https://arxiv.org/abs/1309.0697}{[arXiv:1309.0697 [hep-th]]}.
}

\lref\DonagiMHD{
  R.~Donagi and E.~Sharpe,
  ``On the global moduli of Calabi-Yau threefolds,''
\href{https://arxiv.org/abs/1707.05322}{[arXiv:1707.05322 [math.AG]]}.
}

\lref\CordovaCVG{
  C.~C\'ordova, T.~T.~Dumitrescu and K.~Intriligator,
  ``Exploring 2-Group Global Symmetries,''
\href{https://arxiv.org/abs/1802.04790}{[arXiv:1802.04790 [hep-th]]}.
}

\lref\PantevNZE{
  T.~Pantev and E.~Sharpe,
  ``Duality group actions on fermions,''
JHEP {\bf 1611}, 171 (2016).
\href{https://arxiv.org/abs/1609.00011}{[arXiv:1609.00011 [hep-th]]}.
}

\lref\HarveyRKO{
  J.~A.~Harvey and G.~W.~Moore,
  ``An Uplifting Discussion of T-Duality,''
\href{https://arxiv.org/abs/1707.08888}{[arXiv:1707.08888 [hep-th]]}.
}

\lref\KasselTuraev{
 C.~Kassel and V.~Turaev,
 ``Braid Groups'',
Graduate Texts in Mathematics 247, Springer, 2008.
}

\lref\KapustinDXA{
  A.~Kapustin, R.~Thorngren, A.~Turzillo and Z.~Wang,
  ``Fermionic Symmetry Protected Topological Phases and Cobordisms,''
JHEP {\bf 1512}, 052 (2015), [JHEP {\bf 1512}, 052 (2015)].
\href{https://arxiv.org/abs/1406.7329}{[arXiv:1406.7329 [cond-mat.str-el]]}.
}

\lref\FreedRQQ{
  D.~S.~Freed and M.~J.~Hopkins,
  ``Reflection positivity and invertible topological phases,''
\href{https://arxiv.org/abs/1604.06527}{[arXiv:1604.06527 [hep-th]]}.
}

\lref\YonekuraUFJ{
  K.~Yonekura,
  ``On the cobordism classification of symmetry protected topological phases,''
[arXiv:1803.10796 [hep-th]].
}

\lref\AtiyahFraming{
M.~Atiyah,
``On framings of 3-manifolds,''
Topology {\bf 29}, 1--7, 1990.
}

\lref\Harer{
J.~Harer,
``The Second Homology Group of the Mapping Class Group of an Orientable Surface,''
Invent.~Math. {\bf 72}, 221--239, 1983.
}

\lref\MasbaumRoberts{
G.~Masbaum and J.~D.~Roberts,
``On central extensions of mapping class groups,''
Math.~Ann. {\bf 302}, 131--150, 1995.
}

\lref\FreedHV{
  D.~S.~Freed,
  ``On Determinant Line Bundles,''
Conf.\ Proc.\ C {\bf 8607214}, 189 (1986).
}

\lref\YonekuraWUC{
  K.~Yonekura,
  ``Dai-Freed theorem and topological phases of matter,''
JHEP {\bf 1609}, 022 (2016).
\href{https://arxiv.org/abs/1607.01873}{[arXiv:1607.01873 [hep-th]]}.
}

\lref\VafaXG{
  C.~Vafa and E.~Witten,
  ``Parity Conservation in QCD,''
Phys.\ Rev.\ Lett.\  {\bf 53}, 535 (1984).
}

\lref\WittenABA{
  E.~Witten,
  ``Fermion Path Integrals And Topological Phases,''
Rev.\ Mod.\ Phys.\  {\bf 88}, no. 3, 035001 (2016).
\href{https://arxiv.org/abs/1508.04715}{[arXiv:1508.04715 [cond-mat.mes-hall]]}.
}

\lref\WittenEY{
  E.~Witten,
  ``Dyons of Charge $e\theta/2\pi$,''
Phys.\ Lett.\  {\bf 86B}, 283 (1979).
}

\lref\FreedIUA{
  D.~S.~Freed,
  ``Anomalies and Invertible Field Theories,''
Proc.\ Symp.\ Pure Math.\  {\bf 88}, 25 (2014).
\href{https://arxiv.org/abs/1404.7224}{[arXiv:1404.7224 [hep-th]]}.
}

\lref\WittenAOA{
  E.~Witten,
  ``Three Lectures On Topological Phases Of Matter,''
Riv.\ Nuovo Cim.\  {\bf 39}, no. 7, 313 (2016).
\href{https://arxiv.org/abs/1510.07698}{[arXiv:1510.07698 [cond-mat.mes-hall]]}.
}

\lref\WittenHC{
  E.~Witten,
  ``Five-brane effective action in M theory,''
J.\ Geom.\ Phys.\  {\bf 22}, 103 (1997).
\href{https://arxiv.org/abs/hep-th/9610234}{[hep-th/9610234]}.
}

\lref\AtiyahSignature{
M.~F.~Atiyah,
\href{https://www.google.com/?q=atiyah+signature}{``The signature of fibre-bundles,''}
pp.~73--84,
in {\sl Global Analysis, papers in honor of K.~Kodaira,}
Univ.~Tokyo Press,
1969.
}

\lref\Humphreys{
J.~E.~Humphreys,
``Variations on Milnor's computation of $K_2\Z$,''
\href{https://doi.org/10.1007/BFb0073729}{pp.~304--307,}
in  {\sl ``Classical'' Algebraic K-theory, and connections with Arithmetic.}
Lecture Notes in Mathematics, vol. 342,
Springer, 1973.
}

\lref\Stein{
M.~R.~Stein,
``Generators, Relations and Coverings of Chevalley Groups over Commutative Rings'',
\href{http://www.jstor.org/stable/2373742}{American J.~of Math.\ {\bf 93}, 965 (1971).}
}

\lref\KorkmazStipsicz{
M.~Korkmaz and A.~I.~Stipsicz,
``The second homology groups of mapping class groups of orientable surfaces,''
\href{http://doi.org/10.1017/S0305004102006461}{Math.~Proc.~Camb.~Phil.~Soc.~{\bf 134} 479 (2003).}
}

\lref\Sakasai{
T.~Sakasai,
``Lagrangian mapping class groups from a group homological point of view,''
Algebraic \& Geometric Topology {\bf 12} 267 (2012)
\href{https://arxiv.org/abs/0910.5262}{[arXiv:0910.5262 [math.GT]]}
}

\lref\ClossetBSE{
  C.~Closset, H.~Kim and B.~Willett,
  ``$ {\cal N} = 1$ supersymmetric indices and the four-dimensional A-model,''
JHEP {\bf 1708}, 090 (2017).
\href{https://arxiv.org/abs/1707.05774}{[arXiv:1707.05774 [hep-th]]}.
}

\lref\HarveyIR{
  J.~A.~Harvey and G.~W.~Moore,
  ``Five-brane instantons and $R^2$ couplings in N=4 string theory,''
Phys.\ Rev.\ D {\bf 57}, 2323 (1998).
\href{https://arxiv.org/abs/hep-th/9610237}{[hep-th/9610237]}.
}

\lref\AtiyahRiemannSurface{
M.~F.~Atiyah,
``Riemann surfaces and spin structures,''
\href{http://www.numdam.org/item?id=ASENS_1971_4_4_1_47_0}{Ann.~Sci.~de l'\'E.N.S.~4$^e$ s\'erie, {\bf 4} (1971) 47.}
}

\lref\Pitsch{
W.~Pitsch,
``Un calcul \'el\'ementair de $H_2(\M_{g,1},\Z)$ pour $g\ge 4$,''
C.~R.~Acad.~Sci.~Paris, {\bf 329} (1999) 667.
}

\lref\BachasUM{
  C.~P.~Bachas, P.~Bain and M.~B.~Green,
  ``Curvature terms in D-brane actions and their M theory origin,''
JHEP {\bf 9905}, 011 (1999).
\href{https://arxiv.org/abs/hep-th/9903210}{[hep-th/9903210]}.
}

\lref\VerlindeMZ{
  E.~P.~Verlinde,
  ``Global aspects of electric-magnetic duality,''
Nucl.\ Phys.\ B {\bf 455}, 211 (1995).
\href{https://arxiv.org/abs/hep-th/9506011}{[hep-th/9506011]}.
}

\lref\FriedanUA{
  D.~Friedan and S.~H.~Shenker,
  ``The Analytic Geometry of Two-Dimensional Conformal Field Theory,''
\href{https://doi.org/10.1016/0550-3213(87)90418-4}{Nucl.\ Phys.\ B {\bf 281}, 509 (1987).}
}

\lref\BrownCohomology{
K.~S.~Brown,
``Cohomology of Groups,"
Graduate Texts in Mathematics vol.~87, Springer, 1982
}

\lref\GilkeyBotvinnik{
P.~B.~Gilkey, B.~Botvinnik,
``The eta invariant and the equivariant spin bordism of spherical space form 2 groups,''
in {\sl  New Developments in Differential Geometry}, pp.~213--223,
Mathematics and Its Applications, vol.~350, Kluwer, 1996.
}

\lref\RosenbergII{
J.~Rosenberg,
``$\C^*$ algebras, positive scalar curvature and the Novikov conjecture, II,''
in {\sl Geometric methods in operator algebras}, pp.~341--374,
Pitman Res.~Notes Math.~Ser.~123, Longman, 1986.
}

\lref\Hsieh{
Chang-Tse Hsieh,
``Discrete gauge anomalies revisited,''
to appear.
}

\lref\KirbyTaylor{
R.~C.~Kirby, L.~R.~Taylor,
``Pin structures on low-dimensional manifolds,''
in {\sl Geometry of Low-Dimensional Manifolds,} vol.~2, pp.~177-242,
London Math.~Soc.~Lecture Note Ser.~151, 1990
}

\lref\DasguptaCD{
  K.~Dasgupta, D.~P.~Jatkar and S.~Mukhi,
  ``Gravitational couplings and $\Z_2$ orientifolds,''
Nucl.\ Phys.\ B {\bf 523}, 465 (1998).
[\href{https://arxiv.org/abs/hep-th/9707224}{hep-th/9707224}].
}

\lref\Benson{
D.~Benson, C.~Campagnolo, A.~Ranicki, and C.~Rovi,
``Cohomology of symplectic groups and Meyer's signature theorem,''
[\href{https://arxiv.org/abs/1710.04851}{arXiv:1710.04851 [math.AT]}].
}

\lref\GaberdielUI{
  M.~R.~Gaberdiel and M.~B.~Green,
  ``An $SL(2, \Z)$ anomaly in IIB supergravity and its F theory interpretation,''
JHEP {\bf 9811}, 026 (1998).
[\href{https://arxiv.org/abs/hep-th/9810153}{hep-th/9810153}].
}

\lref\KirbyTaylorPaper{
    R.~C.~Kirby and L~R.~Taylor, L. R.,
    ``A calculation of {${\rm Pin}^+$} bordism groups,''
   \href{https://doi.org/10.1007/BF02566617}{Comment. Math. Helv., 65 (1990) 434--447}.
}

\lref\MoorePC{
  G.~W.~Moore and E.~Witten,
  ``Integration over the $u$-plane in Donaldson theory,''
Adv.\ Theor.\ Math.\ Phys.\  {\bf 1}, 298 (1997).
[\href{https://arxiv.org/abs/hep-th/9709193}{hep-th/9709193}].
}

\noblackbox

\Title{IPMU-18-0044}{\vbox{\centerline{Anomalies of Duality Groups}
\smallskip
\centerline{and }
\smallskip
\centerline{Extended Conformal Manifolds
}}}

\bigskip

\centerline{Nathan Seiberg$^1$,  Yuji Tachikawa$^2$ and Kazuya Yonekura$^2$}

\bigskip

\centerline{$^1$ School of Natural Sciences,
Institute for Advanced Study,}
\centerline{Einstein Drive, Princeton, NJ 08540, USA}
\centerline{$^2$ IPMU, University of Tokyo, Kashiwa, Chiba 277-8583, Japan}

\bigskip
\bigskip
\bigskip

\noindent
A self-duality group $\G$ in quantum field theory can have anomalies.
In that case, the space of ordinary coupling constants $\M$ can be extended to include the space ${\cal F}$ of coefficients of counterterms in background fields.
The extended space $\N$ forms a bundle over $\M$ with fiber ${\cal F}$,
and the topology of the bundle is determined by the anomaly.
For example, the $\G=SL(2,\Z)$ duality of the $4d$ Maxwell theory has an anomaly,
and the space ${\cal F}=\S^1$ for the gravitational theta-angle is nontrivially fibered over ${\cal M}=\H/SL(2,\Z)$.
We will explain a simple method to determine the anomaly when the $4d$ theory is obtained by compactifying a $6d$ theory on a Riemann surface in terms of the anomaly polynomial of the parent $6d$ theory.
Our observations resolve an apparent contradiction associated with the global structure of the K\"ahler potential on the space of exactly marginal couplings of supersymmetric theories.

\bigskip
\Date{March 2018}

\eject

\hbox{}
\vfill
\abstractfont
\listtoc
\writetoc
\immediate\write\tfile{\baselineskip=16pt\noexpand\abstractfont}
\vfill
\hbox{}
\eject

\newsec{Introduction and summary}

\subsec{Background couplings, dynamical couplings and the duality group}

The observables of a quantum field theory depend on dimensionless parameters $\tau$.
A typical example, which we will discuss in detail below, is $4d$ Maxwell theory, where the complex parameter $\tau={\theta\over 2\pi} +{4\pi i\over g^2}$ includes the gauge coupling constant $g$ and the theta-angle $\theta$.
When we limit ourselves to the theory in flat $\R^d$ and to separated-points correlation functions the parameters $\tau$ uniquely specify the theory.

However, the theory also depends on additional choices like the values of contact terms when two operators touch each other.
Similarly, placing the theory in a nontrivial spacetime other than flat $\R^d$ can lead to additional parameters.
One way to think about this is to couple the system to background classical fields and then these additional choices and parameters are coefficients of local counterterms in the background fields.

One such coupling of background fields in four-dimensional field theory is the gravitational theta-parameter
\eqn\thetag{  i\theta_\grav {\sigma\over 16} =i\theta_\grav {1\over 384\pi^2} \int  \tr (R\wedge R) ~,}
where $\sigma$ is the signature of the manifold.
On spin four-manifolds $\theta_\grav$ is $2\pi$-periodic and on non-spin manifold it is $32\pi $-periodic.
(In the rest of this paper we will limit ourselves to spin manifolds.)  Note that this term is an imaginary contribution to the Euclidean space action.
Another such coupling of background fields is the Euler characteristic
\eqn\Eulerc{t_e\chi = t_e{1\over 32\pi^2} \int \epsilon_{ac}^{bd} R^a_b\wedge R^c_d ~.}
Unlike \thetag, the coefficient $t_e$ is not periodic.

Adding terms like \thetag\ and \Eulerc\ as well as other non-topological terms to the action does not affect the separated-points correlation functions in flat spacetime, and therefore such terms are often ignored
in studies on purely quantum field theoretical phenomena.\foot{%
Of course there are studies on quantum field theoretical phenomena where such terms are carefully treated, e.g.~\refs{\WittenGF,\MoorePC}.
Also, such terms naturally arise when quantum field theories are constructed within string theory and/or M-theory, where a function of supergravity fields plays the role of $\theta_\grav$,
and the action of the duality group of string theory and/or M-theory on them was already studied,  see e.g.~\refs{\DasguptaCD,\BachasUM,\GaberdielUI}.
It would be interesting to work out the relations of these works and the analysis presented in this paper.
}
However, such terms affect contact terms and the value of the partition function in curved spacetime.
Therefore, they can lead to important consequences.

We will denote the space of ordinary coupling constants by $\M$ and the larger space, which includes also the coupling constants of background fields by $\N$.
In the case of a family of conformal field theories $\M$ is known as the conformal manifold and we refer to $\N$ as the extended conformal manifold.
One of the main points of this note is that $\N$ is typically a nontrivial bundle over $\M$, and the topology of the bundle is
determined by the anomalies of the duality group of the theory.

The partition function of the theory $Z$ depends on the ordinary coupling constants and on the coefficients of counterterms of background fields; i.e.\ it is a function on $\N$.
What is more interesting, though, is that it is not a function on $\M$; it is a section of a line bundle on $\M$.
In other words, if we ignore the coupling of background fields, $Z$ is not a single valued function of $\tau$ and we need transition functions as we move from patch to patch on $\M$.
There are two ways to have a standard partition function.
First, we do not ignore the coupling of background fields and then $Z$ is a single valued function on $\N$.
Second, we keep the coupling constants of background fields fixed, but we extend the range of $\tau$ to a multiple cover $\widehat \M$ of $\M$, so that the fibration becomes trivial.

\subsec{The case of $4d$ Maxwell theory}

Since the discussion above has been very general and abstract, we should consider a concrete example.
$4d$ Maxwell theory has a single complex coupling constant $\tau$ taking values in the upper half plane $\H$,
and is invariant under the duality group $\G=SL(2,\Z)$ acting on $\tau$,\foot{
When we consider Maxwell theory on non-spin manifolds or spin manifolds without  a specified spin structure, only a subgroup of $SL(2,\Z)$ is a true duality group,
and the difference manifests itself in the spin of the line operators \refs{\MetlitskiYQA,\SeibergRSG,\SeibergGMD}.
To avoid this subtlety,  we limit ourselves to spin manifolds with a choice of spin structure.
In this case the partition function has full $SL(2,\Z)$ invariance and each line operator appears with both integer and half-integer spins.
}
so that it is parameterized by $\M=\H/SL(2,\Z)$.   More precisely, only $PSL(2,\Z)$ acts on $\H$, but we will continue to denote $\M=\H/SL(2,\Z)$.

Next, we consider also the coupling \thetag.
In this note we are mostly concerned with nontrivial topological properties of partition functions and anomalies.
For that purpose, the phase factor \thetag\ plays the crucial role.
Other terms, such as \Eulerc,
only affect the absolute values of partition functions and have trivial topologies in the parameter space (e.g.\ $t_e$ takes values in $\R$ rather than $\S^1$),\foot{
In general, the topology of the bundle $\N \to \M$ is restricted by the following constraint. Suppose that we have a term in the action of the form $\int s B$,
where $B$ is constructed purely from background fields, e.g.\ $\tr (R\wedge R) $ or $\epsilon_{ac}^{bd} R^a_b\wedge R^c_d $,
and $s$ is its coefficient, e.g.\ $\theta_\grav$ or $t_e$. Trivially, $B$ is the (functional) derivative of the logarithm of the partition function with respect to $s$.  This should not change under duality
and hence the duality transformation can change $s$ only by a shift, $s \to s + F(\tau)$, where $F(\tau)$ is a function of the coupling constants of the theory.
In particular, a transformation like $s \to -s$ is forbidden. If the fiber of the bundle is $\R$, as with $t_e$, the topology of the bundle is trivial. On the other hand,
if the fiber of the bundle is $\S^1$, as with $\theta_\grav$, the topology can be nontrivial if $F(\tau)$ is nontrivial.
}
so we will ignore the coupling \Eulerc.\foot{In particular, almost all the examples in this paper are realized by the compactification of some $6d$ theories
on Riemann surfaces, and the anomaly comes from the gravitational anomaly of the $6d$ theory, which affect the phase factors of partition functions.
The $6d$ theory also suffers from a conformal anomaly.
It should affect the $\chi$ dependence of the effective $4d$ theory.
For simplicity, we will not discuss it here and will focus only on the phase of the partition function.
In supersymmetric theories, as in Sec.~5 below, the terms \thetag\ and \Eulerc\ are related by supersymmetry and then the effect of \Eulerc\ is easily determined.}

Before taking into account the duality group, Maxwell theory is parameterized by
\eqn\NcM{
(\tau,\theta_\grav)\in \H \times \S^1~.}
In particular, the coupling $\tau$ takes values in the covering space $\widehat{\M} = \H$.

The duality of the Maxwell theory on closed 4-manifolds was studied in \WittenGF.
As we will see in more detail in the coming sections, we can rephrase the result of \WittenGF\ as follows.
There exists an anomaly of the duality group $SL(2,\Z)$. This anomaly can be compensated by shifting the gravitational theta angle.
In other words, $SL(2,\Z)$ acts on both $\tau \in \H$ and $\theta_\grav \in \S^1$.
Correspondingly, the space $\N$ of the pair $(\tau, \theta)$ forms a nontrivial fiber bundle of the parameter spaces
\eqn\MNMax{\S^1 \to \N \to \M ~,}
where
\eqn\MNMaxe{
\N=(\H\times \S^1)/ SL(2,\Z)   \ ,\qquad
\M=\H / SL(2,\Z)~.}
Only $PSL(2,\Z)$ acts in the second quotient.  However, the center $C \in SL(2,\Z)$ can act nontrivially on $\S^1$ if $SL(2,\Z)$ has an anomaly.

As we will review below, the partition function of the theory is not a single-valued function on $\M$  \WittenGF, but it is a single-valued function on $\N$.
In other words,
we will make a finer distinction among QFTs by declaring that two theories differing by the gravitational theta-angle
(which affects only the partition function on curved backgrounds) are actually different.  Then, the space $\S^1$ of the gravitational theta-angle is fibered nontrivially over the space $\H/SL(2,\Z)$ of $\tau$ and the partition function is a well defined function on $\N$.

In this paper, we will see that this phenomenon is ubiquitous:
the parameter space of background couplings often forms a non-trivial fiber bundle $\N$
over the space $\M$ of dynamical couplings by the effect of anomalies.

\subsec{Other examples}

It is often the case that the partition function is a function on some space $\widehat{\M}$, but it is a section of a line bundle on $\M=\widehat{\M}/\G$, where
$\G$ is an anomalous group.
This fact is not restricted to duality groups and applies to other groups such as the group of gauge transformations.

To illustrate the point, consider the well known situation of a $2d$ chiral fermion coupled to a background $U(1)$ gauge field.
This theory has the standard perturbative $U(1)$ anomaly.  (It also has a gravitational anomaly, that we are going to ignore.)  The anomaly is the statement that the partition function $Z(A)$ is not gauge invariant
\eqn\towdimchiral{
Z(A + d\alpha) =Z(A)  \phi_A (e^{i \alpha})}
with $ \phi_A (e^{i \alpha}) $ depending on $A$ and $e^{i\alpha}$.  This means that $Z(A)$ is a function of $A$ in the space $\widehat \M$ of $U(1)$ gauge fields before gauge identifications , but it is a section on $\M=\widehat{\M}/\G$, where
$\G$ is the space of gauge transformations.
The line bundle on which the partition function takes values is given as $\L = (\C \times \widehat{\M}) /\G$
where the group $\G$ acts on $( \ell, A)  \in \C \times \widehat{\M} $
as $( \ell, A)  \mapsto ( \phi_A (e^{i \alpha}) \ell ,   A+d \alpha  )$.

Instead of having this section we can add to the system a $3d$ bulk, whose boundary is our $2d$ spacetime and replace $Z(A)$ by $\widetilde Z(A)=Z(A) e^{i{\rm CS}(A)}$ with ${\rm CS}(A)$ the Chern-Simons term in the bulk.  Now the partition function $\widetilde Z(A)$ is gauge invariant, but it depends on additional data -- the value of $A$ (modulo gauge transformations) in the bulk.  Using the notation above, we can also say that $\widetilde Z(A)$ is a single valued function on the space of $3d$ gauge fields modulo gauge transformations, $\N$.

The description in terms of the line bundle is general. However,
in many (although not all) cases,
the anomalies are realized as the shifts of counterterms of background fields under the group $\G$.
We denote the space of counterterms by ${\cal F}$. The extended space is given as $\N=({\cal F} \times \widehat{\M})/\G$.
Let us see examples.

For example, in $2d$, a compactified free boson $\varphi\sim \varphi+2\pi$ has the action
\eqn\twods{S={L^2\over 4\pi}\int d^2x \sqrt{g}\partial_\mu \varphi\partial^\mu \varphi + {t_e\over 4\pi} \int d^2 x \sqrt{g} R}
specified by the radius of the circle $L \in \R:=\widehat{\M}$ and the ``dilaton'' $t_e \in \R :={\cal F}$.
($R$ is the $2d$ curvature and the coupling $t_e$ is the $2d$ version of \Eulerc.)
Then T-duality relates $(L,\ t_e)\sim (L'={1\over L},\ t_e'=t_e-\log L)$; i.e.\ it shifts the dilaton.\foot{
In general, if we consider $(n-1)$-form gauge field $a$ in $2n$ dimensions with the action $(K/2 \cdot 2\pi) \int da \wedge * da$,
the argument of \WittenGF\  says that the partition function transforms as $Z(K^{-1}) = K^{(-1)^n\chi/2 }Z(K)$.
In $2d$, $K =L^2$ gives the T-duality and in 4d, $K = {\rm Im}(\tau)$ gives the case of the Maxwell theory when $\theta=0$.
}
However, this does not lead to an anomaly in the duality group, because the effect can be removed by adding a local counterterm ${\log L\over 8\pi}\int d^2x\sqrt g R$ and then $(L,\ t_e)\sim (L'={1\over L},\ t_e)$.

A particularly simple example, which was analyzed in detail in \GaiottoYUP, involves a charged particle on a ring.
Here  $\theta \in \R :=\widehat{\M}$ represents magnetic flux through the ring.
A background $U(1)$ gauge field $A$ couples to its global $U(1)$ symmetry that shifts the coordinate on the ring and its counterterm is $ik\int A$ with integer $k \in \Z : = {\cal F}$.
In this example, we have the identification $(\theta,\ k) \sim (\theta+2\pi,\ k+1)$ which is realized by
a group $\G=\Z$ whose generator acts as $(\theta,\ k) \mapsto (\theta+2\pi,\ k+1)$. The extended parameter space $\N=({\cal F} \times \widehat{\M})/\G = (\Z \times \R)/\Z$
can be parameterized either by $\theta \in \R$,
or by a $\Z$ bundle (parameterized by $k$) over the base $\M$ parameterized by $\theta\in \S^1$.

In Appendix B we will discuss the example of $4d$ $SU(N)$ gauge theories with a $\theta$ parameter.
Here $\M$ is $\S^1$ with $\theta \sim \theta +2\pi$.
The system can be coupled to a background 2-form $\Z_N$ gauge field $B$ with a counterterm proportional to $BB$  \refs{\KapustinGUA,\GaiottoKFA} with an integer coefficient $p$, such that $p \in \Z$ and $p \sim p+N$
(on a spin manifold).  Thus the space of counterterms in a spin manifold is ${\cal F}=\Z_N$.
The proper periodicity is $(\theta,\ p) \sim (\theta+2\pi,\ p +1)$ and therefore, the extended parameter space $\N$ is labeled by $(\theta,\ p) $ with this identification, or by $\theta$ with an extended range $2\pi N$.
An interesting question that this leads to is what happens when $\theta$ is a dynamical field.
Does it take values in $\M$ or in $\N$?  We will answer this question in Appendix B.

Another notable example of our discussion is the theory of a free $2d$ Majorana fermion with mass $m \in \R :=\widehat{\M}$ (not summed over its spin structures).
For zero mass the theory has a $\Z_2:=\G$ global symmetry, which flips the sign of the left-moving fermion.
The mass term is odd under this symmetry.
In the Ising model description of the system this $\Z_2$ global symmetry is related to Kramers-Wannier duality.
In the language of this paper the parameter space $\M$ is parameterized by $m\ge 0$; i.e.\ it is obtained by modding out $\widehat{\M}=\R$ by the $\G=\Z_2$ symmetry.
As emphasized in \KapustinGUA, the system also has a discrete gravitational theta-parameter $i\zeta \Arf(\Sigma)$, where $\zeta\in\{0, \pi\}$ and $\Arf(\Sigma)$ is the Arf invariant of the spin structure of the underlying spacetime $\Sigma$. The space of counterterms is ${\cal F}=\{0, \pi \}$.
The theory is left invariant only if we combine the change of sign of $m$ with a shift of $\zeta$ by $\pi$.
As in all our examples, this can be viewed as a mixed anomaly between Kramers-Wannier duality and gravity.
Clearly, the extended parameter space $\N$ can be parameterized either by $m\ge 0$ and $\zeta\in\{0, \pi\}$, or by an arbitrary real $m$.

\subsec{Outline of the paper}

The rest of the paper is organized as follows:

In section 2, we will discuss in detail the anomaly of the duality group of $4d$ $U(1)$ Maxwell theory,  expanding the brief summary we gave in section 1.2 above.
We first remind ourselves how $SL(2,\Z)$ acts on the operators of the theory,
and then recall the anomalous transformation of the partition function first pointed out in \WittenGF.
We will see that no counterterms can remove this anomalous transformation.
This means that the partition function is a section of a nontrivial line bundle on $\M=\H/SL(2,\Z)$.
We then explain that we can regard the partition function as a genuine function, by incorporating the gravitational theta angle $\theta_\grav$.

In section 3, to prepare for the generalization in the latter part of the paper,
we systematically study how we can analyze the anomaly of the duality group,
in terms of the line bundles over the configuration space of background fields.
We will find it particularly useful to compare the anomaly of the duality group
with the anomaly of a $2d$ conformal field theory due to the chiral central charge,
which we will explain at length.

In section 4, we extend our discussion of a $4d$ $U(1)$ Maxwell theory to $4d$ $U(1)^g$ theory.
This will be done by considering this $4d$ theory as arising from the compactification of the $6d$ self-dual tensor field on a Riemann surface $\Sigma_g$ of genus $g$.
We then further generalize our consideration to the compactification of arbitrary $6d$ theory on $\Sigma_g$.
We will see that the anomaly of the duality group in $4d$ can be obtained from the anomaly polynomial of the $6d$ theory.

In section 5, we consider generic $4d$ \Nequals2 superconformal theories, that do not necessarily arise from a $6d$ theory.
Exactly marginal couplings of such a theory parameterize a K\"ahler manifold $\M$ known as the conformal manifold of the theory.
Adding the space of background couplings we have a fiber bundle $\N$ over $\M$.
We will use the results of \GomisYAA\ to relate the K\"ahler class of $\M$ to the bundle structure of $\N$,
and show that the K\"ahler potential is globally well defined on $\N$.
This resolves an apparent conflict between the claims in \GomisYAA\ and the discussion in \DonagiVWH.

We have two appendices:
In Appendix A, we give a brief discussion of $5d$ symmetry protected topological phases that capture the anomaly of the $4d$ duality group we discuss in the main text.
In Appendix B, we explain that many of the phenomena discussed in this paper also arise in the context of  $4d$ $SU(N)$ gauge theory, where the $\theta$ angle of the $SU(N)$ theory parameterizes $\M$,
the shift of $\theta$ by $2\pi$ plays the role of the duality group, and a certain coupling of the background field for the $\Z_N$ one-form symmetry extends $\M$.

\newsec{Anomaly of duality group: Demonstration in free Maxwell theory }

In this section, we demonstrate the existence of anomalies of duality groups.
We will consider the free Maxwell theory on spin manifolds,
\eqn\MaxwellL{
S =  {1 \over g^2} \int F \wedge \star F  - {i \theta \over 8\pi^2} \int F \wedge F
}
as an example. On spin manifolds, this theory has the $SL(2,\Z)$ duality group.
The transformation of partition functions under the $SL(2,\Z)$ was discussed in \WittenGF.
We use the result there to see that there is a mixed 't~Hooft anomaly between background gravity and $SL(2,\Z)$, which cannot be canceled by local counterterms.

\subsec{Generalities}

The duality group $SL(2,\Z)$ is quite analogous to the diffeomorphism group and the gauge transformation group.
Let us consider the diffeomorphism group as an example.
The diffeomorphism group acts on the background metric $g_{\mu\nu}$ (as well as other fields). In the same way, $SL(2,\Z)$
acts on the exactly marginal coupling $\tau = 4\pi i/g^2 + \theta/2\pi$ regarded as a background field.
For special configurations of $g_{\mu\nu}$, such as the flat metric $g_{\mu\nu}=\eta_{\mu\nu}$, AdS metric, etc.,
there is a subgroup of the diffeomorphism group that fixes that particular metric, e.g.\ the Poincar\'e group for the flat metric and
$SO(2,d-1)$ for the AdS metric. They are the symmetry group for that particular metric.
In the same way, for special values of $\tau$, such as $\tau=i$ and $\tau=e^{\pi i /3}$,
there is a subgroup of $SL(2,\Z)$ that fixes that particular value of $\tau$, and that subgroup can be regarded as a symmetry group of the theory
for that value of $\tau$.\foot{More precisely, an extension of this subgroup is part of the symmetry group of the theory at this value of $\tau$.}
The analogy between the diffeomorphism group and the $SL(2,\Z)$ duality group becomes even more transparent when $SL(2,\Z)$ originates from a compactification of a higher dimensional theory on a torus, as we will discuss in a later section.

By gluing local patches of spacetime using local diffeomorphism groups, we get manifolds with nontrivial topology.
In the same way, by gluing local patches using $SL(2,\Z)$, we can get a nontrivial bundle of $SL(2,\Z)$ in spacetime.
However, we restrict our attention to the case that the $SL(2,\Z)$ bundle is trivial for simplicity.

In general, let $\Phi$ be a set of background fields, such as $\Phi =\{g_{\mu\nu}, \tau, \cdots \}$.
Also let $\G$ be a group that acts on the background fields as well as the dynamical fields (or operators) of the theory.
Examples of the group $\G$ include the diffeomorphism group, $SL(2,\Z)$, and many others.

In our discussions in this paper, we take the duality group to be the one that acts properly on the operators of the theory in the following sense.
For example, the background field $\tau$ itself is acted by $PSL(2,\Z)$,
but $PSL(2,\Z)$ does not act properly on the operators, because the center $C=S^2 \in SL(2,\Z)$
acts nontrivially on the fields as $(\vec{E}, \vec{B}) \to (-\vec{E}, -\vec{B})$.
Indeed, the pair $(\vec{E}, \vec{B})$ in an appropriate normalization transforms in the two dimensional defining representation of $SL(2,\Z)$.
The transformation under $T \in SL(2,\Z)$ is $(\vec{E}, \vec{B}) \to (\vec{E}+\vec{B}, \vec{B})$, which is the Witten effect \WittenEY,
while the transformation under $S \in SL(2,\Z)$ is the standard electric-magnetic duality transformation $(\vec{E}, \vec{B}) \to (\vec{B}, -\vec{E})$.
So we must use $SL(2,\Z)$ instead of $PSL(2,\Z)$ for the Maxwell theory.
Depending on the theory, we may have to consider a further covering group.
For example, in the case of a $4d$ free fermion that arises from the compactification of a $6d$ Weyl fermion on a torus,
we may consider a version of the spin cover of $SL(2,\Z)$.
Another example is the T-duality of a $2d$ compact scalar theory.  It acts as $\Z_2$ on the exactly marginal operators, but at the self-dual radius it acts as a $\Z_4$ symmetry; its generator squares to minus one on the operators with half-integer $SU(2)$ quantum numbers (see e.g.\ \refs{\PantevNZE,\HarveyRKO} for recent discussions).

We denote the configuration space of background fields $\Phi$ before dividing by $\G$ as $\widehat{\M}$.
For example, in the case of $\tau$ which depends on spacetime, it is
$
\widehat{\M}  = {\rm Map}(X, \H),
$
where $\H=\{ z \in \C;\ {\rm Im}(z) >0 \}$ is the upper half plane, $X$ is the spacetime manifold, and ${\rm Map}(X, \H)$ means the space of smooth maps
from $X$ to $\H$.
If we restrict attention to constant $\tau$, $\widehat\M$ is reduced to $\H$.
Thus, in the following, we often write
$
\widehat{\M}= \H
$
for constant $\tau$.

Now, if the group $\G$ is anomaly free, it makes perfect sense to say that we consider the configuration space as
\eqn\modspace{
\M : = \widehat{\M}/\G,}
and say that the partition function is really a function on this quotient space.
For example, partition functions on nontrivial manifolds are usually considered as a function of the metric $g_{\mu\nu}$ up to diffeomorphism
transformations. However, when there is an anomaly of the diffeomorphism group, the partition functions are no longer invariant under
diffeomorphisms. If the non-invariance of the partition functions cannot be cancelled by local counterterms, that represents a true anomaly.
Below we will see that $SL(2,\Z)$ also has such an anomaly in the free Maxwell theory.

\subsec{Anomaly of $SL(2,\Z)$}

  The $SL(2,\Z)$ duality group of the free Maxwell theory has the following anomaly.
We take $\tau$ to be constant in spacetime.
The generators $S$ and $T$ of $SL(2,\Z)$ act on $\tau$ as $T: \tau \to \tau+1$ and $S: \tau \to -1/\tau$.
The partition function on closed, oriented, spin four-manifolds transforms as~\WittenGF\
\eqn\uonetrans{\eqalign{
&Z(\tau+1, \bar \tau+1)=Z(\tau,\bar \tau)\cr
& Z(-1/\tau, -1/\bar\tau)=\tau^{(\chi+\sigma)/4}   \bar\tau^{(\chi-\sigma)/4} Z(\tau,\bar\tau)~,}}
where $\chi$ and $\sigma$ are the Euler number and the signature of the manifold, respectively.\foot{On a nonspin manifold more care is needed.  First, there can also be a coupling of the form $\pi i \int c_1 \cup w_2$, where $c_1$ is the first Chern class of the Maxwell $U(1)$ bundle and $w_2$
is the second Stiefel-Whitney class of the manifold. Second, for ${\rm Re}(\tau)=\theta=0\ {\rm mod}\ 2\pi$ the partition function is manifestly real and positive if the above coupling is also absent.
 In order for the $S$ transformation to preserve this fact, we need to add a factor of $(-i)^{\sigma/2}$ in its action in \uonetrans.  When the theory is on a spin manifold this factor is trivial because $\sigma$ is a multiple of 16.}

Let us start the analysis of the anomaly.
First, we note that the partition function is not invariant under the $S$ transformation $\tau \to -1/\tau$ for generic values of $\tau$.
To make sure that this represents a true anomaly, we have to consider all possible local counterterms in the Lagrangian.
Since the above transformation depends on the Euler number and the signature, we can restrict our attention to counterterms of the form
\eqn\counterterm{
 \L_{\rm counterterm} = f(\tau, \bar\tau)({\rm Euler~density}) + g(\tau, \bar\tau)({\rm Signature~density})  }
where $f(\tau, \bar\tau)$ and $g(\tau, \bar\tau)$ are arbitrary, not necessarily holomorphic functions of $\tau$ satisfying
certain reality conditions consistent with unitarity. Notice that they are local in spacetime when we extend $\tau$ to a spacetime-dependent field $\tau(x)$.

A simple argument shows that no choice of counterterms \counterterm\ can cancel the anomaly in \uonetrans.  Consider the partition function at $\tau = e^{\pi i/3}$, which is fixed by $ST^{-1}$.
Using \uonetrans\ we see that $ST^{-1}$ leads to a nontrivial phase factor $e^{\pi i \sigma/3}$. (It is nontrivial even on a spin manifold where $\sigma\in 16\Z$.) Since $\tau = e^{\pi i/3}$ is fixed by this transformation, counterterms like \counterterm\ do not transform and cannot cancel this phase.

The theory at $\tau= e^{\pi i/3}$ has a global $\Z_3$ symmetry generated by $ST^{-1}$, or an extension of it.  Writing $\tau = e^{\pi i/3} +\delta \tau$ with infinitesimal $\delta \tau$ the action $ST^{-1}(e^{\pi i/3} + \delta \tau) =-{1\over e^{\pi i/3}-1 + \delta \tau }\approx e^{\pi i/3} +e^{2\pi i/3}\delta\tau$ shows that $\delta \tau $ transforms under this $\Z_3$.  Equation \uonetrans\ means that whenever $e^{\pi i \sigma/3}\not=1$ there is a mixed anomaly between this global $\Z_3$ symmetry and gravity.  $\delta\tau$ couples to an exactly marginal operator $O$ and the anomaly means that $Z(\tau = e^{\pi i/3})=\langle 1\rangle_{\tau = e^{\pi i/3}}$ is nonzero only when  $e^{\pi i \sigma/3}=1$, $\langle O\rangle_{\tau = e^{\pi i/3}}$ is nonzero only when  $e^{\pi i \sigma/3}=e^{2\pi i /3}$, and $\langle \bar O\rangle_{\tau = e^{\pi i/3}}$ is nonzero only when  $e^{\pi i \sigma/3}=e^{-2\pi i /3}$.\foot{This is analogous to the following situation.
Let us consider a free Weyl fermion in a nontrivial gravity background with nonzero $\sigma$.  This system has
$-\sigma/8$ net fermion zero modes, which saturate a mixed $U(1)({\rm grav})^2$ anomaly.  These zero modes constrain the nonzero amplitudes to have some number of fermion insertions that depends on $\sigma$.}

We can do the same analysis for the generator $S \in SL(2,\Z)$ by taking $\tau = i$, which is fixed by $S$.
Then we get a trivial phase factor there.
Thus, the transformation $S$ is anomaly free on spin manifolds, at least at the point $\tau = i$.

The structure of the anomaly on the entire $\H$ plane can be simplified by choosing an appropriate counterterm.
For this purpose, we use the Dedekind eta function $\eta(\tau)$,
which is nonzero on the entire upper half-plane $\H$.
For Maxwell theory, we choose the local counterterm \counterterm\ as
$f(\tau,\bar\tau) = {\rm Re} \log \eta(\tau) $ and $g(\tau, \bar\tau) = i {\rm Im} \log \eta(\tau) $.
Notice that the logarithm of $\eta$ makes sense because $\eta$ is everywhere nonzero.
This counterterm modifies the partition function $Z(\tau, \bar\tau)$ of the original Maxwell theory to
\eqn\addcounter{
Z'(\tau, \bar\tau) : = \eta(\tau)^{-(\chi+\sigma)/2}   \eta( -\bar{ \tau} )^{ -(\chi-\sigma)/2} Z(\tau, \bar\tau).
}
By using $\eta(\tau+1)=e^{\pi i /12}\eta(\tau)$ and $\eta(-1/\tau) = \sqrt{- i \tau} \eta(\tau)$,
we see that the new partition function after adding the counterterm transforms as
\eqn\newtrans{\eqalign{
&Z'(\tau+1, \bar \tau+1)= e^{-\pi i \sigma /3} Z'(\tau,\bar \tau)\cr
& Z'(-1/\tau, -1/\bar\tau)=Z'(\tau,\bar\tau)~.}}
Here we used $e^{-\pi i \sigma /12} = e^{-\pi i \sigma /3}$, which is valid for $\sigma$ a multiple of $16$.
The counterterm completely cancels the anomaly of the $S$ transformation, while the $T$ transformation
now has an anomaly $e^{-\pi i \sigma /3}$, which is a $\tau$-independent phase factor.  As a check, above we found an anomaly $e^{\pi i \sigma /3}$ in $ST^{-1}$ at the point $\tau = e^{\pi i/3}$.  This can be derived more easily using \newtrans\ and now this anomaly is true for all $\tau$ and not only at  $\tau = e^{\pi i/3}$ .

In particular, we can see that three copies of Maxwell theory coupled to the same $\tau$
is anomaly free, meaning that $(Z'(\tau, \bar\tau) )^3$ is invariant under $SL(2,\Z)$, at least if $\tau$ is constant in spacetime.

The anomaly found above is interpreted as a mixed anomaly between $SL(2,\Z)$ and gravity, because
it depends on the signature $\sigma$.
It would be interesting to study whether there is any pure $SL(2,\Z)$ anomaly, which is independent of the signature $\sigma$,
analogous to the anomaly $U(1)^3$ in the Weyl fermion theory.
To detect such an anomaly requires us to consider nontrivial $SL(2,\Z)$ bundles.
We will not attempt to do that in this paper, but our computation in Appendix A implies that three copies of Maxwell theory has a pure $SL(2,\Z)$ anomaly.

\subsec{Extension of the space of couplings by the gravitational theta angle}
Because of the anomaly found above, the partition function is not a function on $\M=\H/SL(2,\Z)$.
However, we can still regard it as a function on an $\S^1$ bundle over $\M$ as we now explain.

For this purpose, we introduce the gravitational theta angle $\theta_\grav$ defined in \thetag\ as part of the background fields
$\Phi=\{g_{\mu\nu}, \tau, \theta_\grav \}$.
We regard it as taking values in $\S^1=\R/2\pi \Z $ by the identification $\theta_\grav \sim \theta_\grav + 2\pi$.

We also need the following one-dimensional representation of $SL(2,\Z)$ which plays an important role also in the next section.
First, recall that $SL(2,\Z)$ can be represented as
\eqn\SLtowZ{
SL(2,\Z)=\vev{S,T \mid S^2=(T^{-1}S)^3,\ S^4=1}.
}
Then, a one-dimensional (and hence abelian) representation $\varphi$ of $SL(2,\Z)$ must satisfy $\varphi(S)=\varphi(T)^3$ and $\varphi(T)^{12}=1$.
Therefore, a general one-dimensional representation is given as
\eqn\rep{
\varphi_n(T)=e^{-2\pi i n/12}, \qquad \varphi_n(S)=e^{-2\pi i n/4},
}
where $n \in \Z_{12}$ is an integer modulo 12.   Here we will be particularly interested in the special representation with $n=8$
\eqn\repeight{
\varphi_8(T)=e^{2\pi i/ 3}, \qquad \varphi_8(S)=1.}

We define the action of $g \in SL(2,\Z)$ on the pair $(\tau, \theta_\grav) \in \H \times \S^1$ as
\eqn\combinedST{
 (\tau, \theta_\grav) \mapsto (g \cdot \tau,\ \theta_\grav + i \log \varphi_{8}(g) ),
}
where $g \cdot \tau$ means the action of $g$ on $\tau$, e.g., $S \cdot \tau = -1/\tau$ and $T \cdot \tau = \tau+1$.

Now the partition function including $\theta_\grav$ is given by
\eqn\thetainclude{
Z'(\tau, \bar\tau, \theta_\grav) : = Z'(\tau, \bar\tau) e^{i  \theta_\grav \sigma / 16}.
}
By using \newtrans\ we can see that
\eqn\newtranstwo{
Z'( g \cdot (\tau, \bar \tau,\theta_\grav) )= Z'(  \tau, \bar \tau,\theta_\grav )~,
}
where $g \cdot (\tau, \bar \tau,\theta_\grav)$ means the action of $g \in SL(2,\Z)$
on $(\tau, \bar \tau,\theta_\grav)$ defined above. Therefore, the partition function is invariant under $SL(2,\Z)$ by including
the transformation of $\theta_\grav$. Then the partition function can be regarded as a function on
\eqn\extbytheta{
\N : = (\H \times \S^1)/SL(2,\Z).
}
This is an $\S^1$ bundle over $\M=\H/SL(2,\Z)$.
This is the extended conformal manifold discussed in the introduction.

\newsec{A systematic treatment of anomalies by line bundles}

As discussed in the previous section, let $\widehat{\M}$ be the configuration space of background fields
such as $\Phi=\{g_{\mu\nu}, \tau,\cdots \}$. Let $\G$ be the group that acts
on those background fields as well as dynamical fields, such as the diffeomorphism group and $SL(2,\Z)$.
We assume that the partition function is a function on $\widehat{\M}$. However,
if the group $\G$ has an anomaly, then the partition function is not a function on $\M=\widehat{\M}/\G$.
Instead, it is a section of a line bundle on $\M$. We use the line bundles to discuss a more systematic approach
to the anomalies of duality groups.

\subsec{Generalities}

More precisely, the relevant line bundles are equivariant line bundles defined as follows.
For simplicity, we assume that $\widehat{\M}$ is topologically trivial, although this assumption can be relaxed.
Then any line bundle on $\widehat{\M}$ is topologically trivial and hence of the form
$\C \times \widehat{\M}$, where $\C$ is the fiber of the line bundle.

Let $(\ell, \m) \in \C \times \widehat{\M}$ be a coordinate system.
Suppose that there is an action of $g \in \G$ on this total space as
\eqn\actionG{
 (\ell, \m) \to (\phi_\m (g) \ell,\ g \cdot \m),
}
where $g \cdot \m$ is the action of $g \in \G$ on $\m \in \widehat{\M}$ and $\phi_\m (g) $ takes values in $\C \setminus \{0\}$.
This $\phi_\m (g) $ must satisfy the relation $\phi_\m (fg)=\phi_{ g \cdot \m} (f) \phi_\m (g)$ so that the group action of $\G$ is well defined.
An equivariant line bundle $\L$ is the bundle $\C \times \widehat{\M}$ with such a specified action of $\G$.

If $\G$ acts freely on $\widehat{\M}$ and hence has no fixed points, then we can consider the space $\L=(\C \times \widehat{\M} )/\G$, which
is really a line bundle over $\M = \widehat{\M}/\G$. Although the action of $\G$ is not free in our application discussed below,
we often loosely call the equivariant line bundle on $\widehat{\M}$ as a line bundle on $\M$.

The equivariant line bundle is relevant to the anomaly for the following reason.
Let $Z(\m)$ be the partition function as a function of $\m \in \widehat{\M}$.
If there is no anomaly, the partition function is invariant under $g \in \G$ as $Z(g \cdot \m) = Z(\m)$.
However, the existence of an anomaly in $\G$ means that the partition function
is not invariant under $g \in \G$, but transforms as
\eqn\ptransform{
Z(g \cdot \m) = \phi_\m (g) Z(\m)
}
for some nontrivial factor $ \phi_\m (g)$, which is a local functional of background fields.\foot{
Actually, there exist more subtle anomalies such that we always have $ \phi_\m (g) = 1$, but still the partition function suffers from a phase ambiguity.
See \WittenABA\ for discussions and examples. Here we neglect those subtle anomalies.
They are believed to be treated in the framework of symmetry protected topological (SPT) phases, or equivalently invertible field theories \FreedIUA.
We give a brief discussion of SPT phases of duality groups in Appendix A.
}
We call a function $Z(\m)$ on $\widehat{\M}$ a section of the equivariant line bundle determined by $ \phi_\m (g)$
if it transforms as in \ptransform .
Therefore, we say that the partition function with an anomaly $ \phi_\m (g)$ takes values
in a line bundle on $\M$.

In summary, a partition function $Z(\m)$ is {\it a function on $\widehat{\M}$}, while it is {\it a section of a line bundle on $\M$}.

Anomalies often mean that the definition of the partition function needs more data.
In the case of anomalies in the diffeomorphism group or local gauge transformation group we extend the gauge fields to higher dimension and the additional data is in this extension.
In the case of anomalies of duality groups, it depends on whether we consider nontrivial $SL(2,\Z)$ bundles on spacetime or not.

If we consider a nontrivial bundle of the duality group on spacetime,
we extend the bundle and the configuration of the couplings to higher dimension as in the case of the diffeomorphism group or local gauge transformation group.
Such a case becomes  important in some applications to the duality group $SL(2,\Z)$ of string theory, such as the orientifold background in type IIB string theory and F-theory.

On the other hand, if we only consider trivial $SL(2,\Z)$ bundles and constant couplings, there is an alternative way to treat the anomaly.
We may extend the space of couplings by including counterterms of the background fields.
Let ${\cal F}$ be the parameter space of those counterterms. Then, a partition function is {\it a function on the extended space
$\N = (\widehat{\M} \times {\cal F})/\G$}, where the action of $\G$ on ${\cal F}$ is determined by the anomaly.  Below we will consider a $4d$ theory that arises from the compactification of a $6d$ theory.
Then the diffeomorphism anomaly of the $6d$ theory becomes the modular anomaly of the $4d$ theory.

\subsec{Line bundles on $\H$}

Now we specialize our attention to the case $\G = SL(2,\Z)$ and $\widehat{\M} =\H$.
We fix a spacetime manifold with signature $\sigma$.

We consider a line bundle $\L$ over $\H/SL(2,\Z)$.
More precisely, it is an equivariant line bundle over $\H$ with an $SL(2,\Z)$ action. Namely, we consider a line bundle
$\C \times \H$ and
specify an action of $SL(2,\Z)$ on this total space.
Let $(\ell, \tau)$ be the coordinate system of $\C \times \H$.
Then the action of $g \in SL(2,\Z)$ is given by $(\ell, \tau) \to ( \phi_\tau(g) \ell, ~g\cdot \tau)$.
$\phi_\tau(g)$ determines the properties of the equivariant line bundle $\L$.
In the following, we loosely say that $\L$ is a line bundle on $\H/SL(2,\Z)$.

We can construct a class of line bundles
by using the one-dimensional representations $\varphi_n$ of $SL(2,\Z)$ defined in \rep.
We take $\phi_\tau(g)$ to be $\varphi_n(g)$, which is independent of $\tau$.
The line bundle corresponding to $\varphi_{n=1}$ is denoted as $L$. This is called the Hodge line bundle.
For general values of $n$, the line bundle
corresponding to $\varphi_n$ is given by $L^{\otimes n}$. The representation $\varphi_n$ is trivial for $n=12$, so $L^{\otimes 12}$ is a trivial bundle.
It turns out that this class of line bundles are enough for $SL(2,\Z)$. The reason behind it will be explained by
using group cohomology of $SL(2,\Z)$ in section 4.2.

\subsec{$2d$ central charge and line bundles}

In section 4, we discuss $4d$ theories that are obtained by compactification of $6d$ theories on Riemann surfaces.
The free Maxwell theory discussed in the previous section is one such example. Then, we will show that
the anomaly of the duality groups of the $4d$ theory is actually related to the gravitational anomaly of the $2d$ theory, which is obtained by
the compactification of the $6d$ theory on $4d$ manifolds with signature $\sigma$.
Therefore, here we discuss the anomalies of $2d$ theories.

In the following discussion we will use the language of $2d$ CFTs, although we do not use conformal invariance.
The gravitational anomaly of a $2d$ CFT is characterized by its chiral central charge $c_L-c_R$, which we will denote by $c$.
In general, the partition function of such a theory on a torus is a section of a vector bundle on $\H/SL(2,\Z)$, whose rank is the number of conformal blocks.
Here we work under a simplifying assumption that the partition function is a section of a \emph{line} bundle.
Equivalently, we assume that there is a unique torus character.
Examples include $b,c$ ghost systems with integer spins and the level-1 chiral $E_8$ theory.

It is known that this line bundle is given by $L^{\otimes(c/2)}$ where $L$ is the line bundle corresponding to $\varphi_{n=1}$ defined in the previous subsection.\foot{%
The detailed explanation on this and other points on the geometry of the moduli space of Riemann surfaces and $2d$ CFT can be found in the classic paper by Friedan and Shenker in \FriedanUA.
}
A quick way to see it is to notice the following facts.
The partition function (or conformal block) of a $2d$ CFT
transforms in a representation of $SL(2,\Z)$.  If there is only one torus character, it transforms as $\varphi_n$ for some $n$.
If the operators of the theory have integer spins and the chiral central charge is $c$, it transforms under $T \in SL(2,\Z)$ as $Z(\tau +1)=e^{-2\pi i c/ 24} Z(\tau)$.

On the other hand, if the partition function $Z(\tau)$ is a section of $L^{\otimes n}$,
then it transforms as $Z(\tau+1)=\varphi_n(T)Z(\tau)$ by the definition of the equivariant line bundle.
We have $\varphi_n(T)=e^{- 2\pi i n/12}$.
By comparing them, we get $c = 2n$.
We conclude that
\eqn\correspondence{
L^{\otimes n}  \quad\longleftrightarrow \quad  \varphi_n  \quad\longleftrightarrow   \quad c=2n.
}

We note here that if we further require that the theory is unitary as a $2d$ theory,
$n$ is required to be a multiple of four, and $c$ is required to be a multiple of eight.
This is because reflection positivity implies that the partition function on a square torus is positive and real,
and therefore $\varphi_n(S)$ needs to be $1$ there.

Let us put Maxwell theory in this general framework.
The transformation \newtrans\ implies that the relevant transformation function is $\varphi_{n=\sigma/2}$
and hence the line bundle is $L^{\otimes (\sigma/2)}$.
Therefore, it corresponds to the case that the chiral central charge is
\eqn\cLcR{
c = \sigma.
}
We will reproduce this result by another method in section 4.4.

The central charges obtained for $SL(2,\Z)$ is meaningful only modulo $24$, since $n=c/2$ is meaningful only modulo $12$.
When we consider a compactification of $6d$ theories on higher genus Riemann surfaces in a later section, the value of $c$ will be meaningful as an integer.

\subsec{Extension of the space of couplings by the gravitational theta angle}

By extending the space of the couplings $\tau \in \H$ by adding the gravitational theta angle $\theta_\grav \in \S^1 = \R/2\pi \Z$,
we can make the partition function a well defined function on $\N = (\H \times \S^1)/SL(2,\Z)$, as in the case of  Maxwell theory.
This is possible if the central charge $c$ is given by the signature $\sigma$ as
\eqn\csigma{
{c \over 2}=n= \kappa { \sigma \over 16}~,
}
where $\kappa$ is an integer, which is independent of the spacetime manifolds.
We will explain the reason behind the proportionality between $c$ and $\sigma$ in a later section. Maxwell theory has $\kappa = 8$.

Then, we impose the transformation law of $\theta_\grav$ under $g \in SL(2,\Z)$ as
\eqn\thetatransf{
g \cdot \theta_\grav = \theta_\grav + i \log \varphi_{\kappa}(g).
}
 $\theta_\grav$ appears in the partition function as $e^{ i \theta_\grav \sigma/16}$,
and it transforms as
\eqn\thetatransftwo{
g \cdot e^{ i \theta_\grav \sigma/16} = \varphi_{n}(g)^{-1}e^{ i \theta_\grav \sigma/16}.
}
where $n$ and $\kappa$ are related as above. From this, it is easy to see that the total partition function
$Z(\tau, \bar{\tau}, \theta_\grav)=Z(\tau, \bar{\tau}) e^{ i \theta_\grav \sigma/16}$ is invariant under $SL(2,\Z)$
and hence it can be regarded as a function on the extended space $\N = (\H \times \S^1)/SL(2,\Z)$.
This is an $\S^1$ bundle over $\M = \H /SL(2,\Z)$: $\S^1 \to \N \to \M$.
The map $\N \to \M$ is given by forgetting the $\theta_\grav$.

\subsec{Remark on the spin structure dependence of the anomaly}

In many situations the duality group arises from compactification of a higher dimensional theory on a torus.
The above result for the anomaly, $n=c/2$, was derived under the assumption that there is no spin structure dependence on the torus.
However, often such a compactification does depend on the choice of spin structure.  A simple example is the free $6d$ Weyl fermion compactified on a torus.

The spin structures on the torus fall into two classes, three even spin structures and a single odd spin structure.
In the even spin structures the spinors have anti-periodic boundary condition around some of the directions;
the direction does not matter because they can be exchanged by $SL(2,\Z)$.
In the odd spin structure the spinors have periodic boundary conditions in all directions.

The odd spin structure is more interesting, because after the compactification on the torus it leads to massless fermions.  Also, it is unique and therefore we expect $SL(2,\Z)$ to act simply.  However, there are some changes from the above discussions.

First, $SL(2,\Z)$ should be replaced by its
spin double cover, defined as
\eqn\Mptzd{Mp(2,\Z):=\vev{S,T \mid S^2=(T^{-1}S)^3,\ S^8=1}~.}
The one dimensional representations of this group are as in \rep, except that now $n$ can also be half-integer.  The basic representation is $\varphi_{1/2}(T)=e^{-2\pi i /24}$ and $\varphi_{1/2}(S)=e^{-2\pi i /8}$.  It corresponds to the line bundle $L^{1/2}$.

Second, unlike the previous discussion about the comparison with two-dimensional conformal field theories, here the action of $T$ is $e^{ - 2\pi i c' /24}$, and again $ n = c'/2$, but now
$-c' /24$ is the ground state energy of the $2d$ CFT in the Ramond-sector and in general it differs from the central extension $c$ of the Virasoro algebra.
For example, a single $2d$ Majorana-Weyl fermion has $c= 1/2$ and $c'=-1$.
Thus a Majorana-Weyl fermion has the smallest possible absolute value of $n$.

\newsec{$6d$ perspective and higher-genus extension}

\subsec{$4d$ Maxwell, $6d$ self-dual tensor and $2d$ chiral bosons}

The anomaly in $SL(2,\Z)$ of $4d$ Maxwell theory can also be understood
from a $6d$ perspective.
A $4d$ $U(1)$ gauge theory results from putting a $6d$ self-dual tensor field on a torus.\foot{
The compactification of a $6d$ self-dual tensor field also gives a compact scalar in $4d$. We will discuss it more at the end of section 4.4.}
We considered this $4d$ theory on a closed 4-manifold $M^{(4)}$.\foot{%
Throughout this section we will remind ourselves the dimension of the manifold by denoting it by a superscript in parentheses.}
So we are considering the partition function of the $6d$ theory on a manifold, which is a product of the torus and the 4-manifold.
We can re-order the compactification and first compactify on the 4-manifold to obtain chiral bosons on the torus.
The difference of the number of the left-moving ones and the right-moving ones is given by the signature $\sigma(M^{(4)})$ of the $4d$ manifold,
thus reproducing the chiral central charge $c$ in \cLcR.  Here we see that the analogy in section 3 between the one-dimensional representations of the modular group or its extension is associated with the physical torus.\foot{%
We mention that the content of this subsection was already noted in \VerlindeMZ\ which was published soon after \WittenGF.}

\subsec{Generalization to $U(1)^g$ and the mapping class group $\Gamma_g$}

Here we generalize the previous discussion from a compactification on a torus to a more general Riemann surface $\Sigma_g^{(2)}$ with genus $g$. This  allows us to extend the analysis to $U(1)^g$ relatively easily.
The duality group  is $Sp(2g,\Z)$.
What is the anomaly?
To answer this, we realize $U(1)^g$ as the compactification of the $6d$ self-dual tensor field on $\Sigma_g^{(2)}$.

To proceed further, we need to review a few mathematical facts.
The moduli space of genus-$g$ Riemann surface is of the form
\eqn\Mgdef{
\M_g = \T_g / \Gamma_g ~,}
where $\T_g$ is the Teichm\"uller space and $\Gamma_g$ is the mapping class group.
In the notation of the previous sections, we have $\widehat{\M}=\T_g$, $\G=\Gamma_g$ and $\M = \M_g$.
The Teichm\"uller space is topologically a ball.
This implies that the orbifold fundamental group of $\M_g$ is $\Gamma_g$ itself
and that all the higher homotopy groups vanish.
This means that $\M_g \sim B\Gamma_g$ for our purpose, where $B\Gamma_g$ is the classifying space of
the group $\Gamma_g$.

The line bundles on $\M_g$ may be classified as follows.
In general, line bundles $\L$ on a space $X$ are classified by the first Chern class $c_1(\L) \in H^2(X, \Z)$.
By identifying $\M_g \sim B\Gamma_g$, the line bundles on $\M_g$ may be classified by $H^2(B\Gamma_g, \Z)$.
This is the group cohomology of $\Gamma_g$.

To understand it better, we consider the following long exact sequence
\eqn\longe{\matrix{
&H^1(B\Gamma_g,\R) &\to
&H^1(B\Gamma_g,U(1)) &\to
&H^2(B\Gamma_g,\Z) &\to
&H^2(B\Gamma_g,\R) \cr
g=1:& 0 &\to & \Z_{12} &\to& \Z_{12} &\to& 0, \cr
g=2:& 0 &\to & \Z_{10} &\to& \Z_{10} &\to& 0, \cr
g\ge 3 : & 0 &\to& 0 &\to& \Z &\to& \R \cr
}}
associated to $0\to \Z\to \R\to U(1)\to 0$.\foot{%
We note here that $H_1(B\Gamma_1,\Z)=\Z_{12}$,
$H_1(B\Gamma_2,\Z)=\Z_{10}$,
$H_1(B\Gamma_{g\ge 3},\Z)=0$, and that
$H_2(B\Gamma_1,\Z)=0$,
$H_2(B\Gamma_2,\Z)=\Z_2$,
$H_2(B\Gamma_3,\Z)=\Z\oplus\Z_2$,
$H_2(B\Gamma_{g\ge4},\Z)=\Z$.
See \refs{\Harer,\MasbaumRoberts} for a general argument for $g\ge 5$,
and \refs{\Pitsch,\KorkmazStipsicz,\Sakasai} for smaller $g$.
}
The cohomology group $H^2(B\Gamma_g,\Z)$ as a line bundle over $\M_g$ is generated by  the Hodge line bundle $L$, which is constructed as follows.
The fiber of $L$ over a point $p\in \M_g$ is the determinant of the space of the holomorphic differentials of the Riemann surface specified by $p$.
This definition of $L$ works uniformly for all $g$.

The relation with the discussions in section 2.2 is as follows.
Let us consider the case $g=1$ in which $\Gamma_{g=1}=SL(2,\Z)$.
In this case, all the elements of $H^2(BSL(2,\Z),\Z)$ are realized as the image of
$H^1(BSL(2,\Z),U(1))$ as shown in the above exact sequence.
This group classifies one-dimensional representations of $SL(2,\Z)$:
there are twelve one-dimensional representations $\varphi_n~(n \in \Z_{12})$ for $g=1$, see \rep.
Therefore, the line bundles in the case $g=1$ are classified by $\varphi_n$ as in the previous section.
The case $g=2$ behaves similarly, where $\Z_{12}$ is replaced by $\Z_{10}$.

For $g\ge 3$, $H^1(B\Gamma_g,U(1))=0$ and therefore there is only the trivial representation.
In this case, the elements of $H^2(B\Gamma_g,\Z)$ map into $H^2(B\Gamma_g,\R)$, and
this is described by the de Rham cohomology group of the moduli space $H^2(\M_g,\R)$.
This is trivial for $g=1$ and $2$, but is nontrivial for $g\ge 3$.

In summary, the line bundle $L$ over $\M_g$ corresponding to the generator of $H^2(B\Gamma_g,\Z)$
is determined by a one-dimensional representation classified by a torsion $\Z_{12}$ when $g=1$,
and a torsion $\Z_{10}$ when $g=2$,
but it has a non-zero de Rham cohomology class in $H^2(\M_g,\R)$ for $g\ge 3$.

Now, an element of the mapping class group $\Gamma_g$  determines a duality of $U(1)^g$ theory,
and therefore there is a homomorphism $\Gamma_g \to Sp(2g,\Z)$.
This homomorphism gives a map between the classifying spaces as $B \Gamma_g \to BSp(2g,\Z)$.
Then we can pull back the line bundles on $BSp(2g,\Z)$ to $B \Gamma_g \sim \M_g$.
It is also known\foot{%
The line of arguments leading to this fact can roughly be given as follows.
The theory of simple Lie groups was generalized by Chevalley and Steinberg
to a theory of  corresponding groups defined over arbitrary rings,
and $Sp(2g,\Z)$ is one example.
As an application of this general theory, the universal cover of $Sp(2g,\Z)$ for $g\ge 4$ was found to be the restriction of the universal cover of $Sp(2g,\R)$ under the natural inclusion \refs{\Stein,\Humphreys}.
This means that $H^2(BSp(2g,\Z),\Z)=\Z$ for $g\ge 4$.
Also, it had been previously shown in \AtiyahSignature\ that the image of the natural homomorphism from $H^2(BSp(2g,\Z),\Z)$ to $H^2(B\Gamma_g,\Z)$ contains the element corresponding to the Hodge bundle for $g\ge 3$.
This element corresponding to the Hodge bundle is known \MasbaumRoberts\ to be the generator of $H^2(B\Gamma_g,\Z)=\Z$.
Therefore we see $H^2(B\Gamma_g,\Z)\simeq H^2(BSp(2g,\Z),\Z)\simeq \Z$ for $g\ge 4$.
Before leaving the footnote, we note that $H_1(BSp(2,\Z),\Z)=\Z_{12}$,
$H_1(BSp(4,\Z),\Z)=\Z_{2}$,
$H_1(BSp(2g,\Z),\Z)=0$ for $g\ge 3$;
$H_2(BSp(2,\Z),\Z)=0$,
$H_2(BSp(2g,\Z),\Z)=\Z\oplus\Z_2$ for $g=2,3$,
$H_2(BSp(2g,\Z),\Z)=\Z$ for $g\ge 4$.
See the Tables at the end of \Benson.
} that for $g\ge 4$,
any line bundles of $BSp(2g,\Z)$ can be distinguished by considering the corresponding line bundles of $B\Gamma_g \sim \M_g$ after the pullback.

Summarizing the mathematical facts reviewed so far,
the anomaly of the duality group $Sp(2g,\Z)$ is fully determined by
the anomaly of $\Gamma_g$ when $g\ge 4$.
In the following we restrict to our analysis to this case of  $g\ge 4$.
The analysis of the intermediate case $g=2,3$ is subtler and we will not carry it out in this paper.

A $2d$ CFT of chiral central charge $c$ with a one-dimensional conformal block naturally determines a line bundle over $\M_g$ such that
the partition function takes values in that line bundle.
We will encounter in this paper only line bundles of this form, determined by the central charge $c$.
In fact,
every line bundle is of this form since the generator $L^{-1}$ (i.e.\ the inverse of $L$) is realized by
the $bc$ ghost system with weight $(h_b, h_c)=(0,1)$.
Since the $bc$ ghost system has $c=-2$, we see that a $2d$ CFT with the central charge $c$ has partition functions
that takes values in $L^{\otimes (c/2)}$ for general genus $g$  \FriedanUA.
Therefore, a line bundle of $BSp(2g,\Z)$ for $g\ge 4$ can be characterized by the same central charge $c$.

Now it is easy to compute the central charge $c$ for the $U(1)^g$ theory.
As in the computation for $g=1$ in section 4.1, it is
\eqn\selfdual{
c=\sigma ~.
}

\subsec{Chiral fermions in $4d$, $6d$, and $2d$}

We can apply the same consideration, starting from a $6d$ chiral fermion.
When compactified on a $2d$ Riemann surface of genus $g$,
it gives rise to a number of $4d$ fermions.
They are parameterized by $\M_g$, the moduli space of genus-$g$ Riemann surfaces $\Sigma_g$.\foot{%
More precisely, we need to specify a spin structure on $\Sigma_g$.
This reduces the duality group from $\Gamma_g$ to a subgroup preserving the spin structure,
and then extends it by a version of its spin cover.
The line bundle is still specified by the effective $2d$ central charge.
As we only care about this central charge, we gloss over this issue in the rest of the section.
The reader should also refer to section 3.5.
}
The fact that these $4d$ fermions came from $6d$ means that
they couple to a natural connection on $\M_g$.
In particular, its partition function on a closed 4-manifold will be a section of a nontrivial line bundle over $\M_g$.

As recalled above, this is specified by the $2d$ central charge $c$.
This can be easily determined; we simply compactify the $6d$ chiral fermion on a closed 4-manifold $M^{(4)}$ to obtain a $2d$ system.  Fermion zero modes on $M^{(4)}$ lead to massless chiral fermions and their number is
\eqn\numzm{
\int \hat A=-\sigma(M^{(4)})/8~.}
Since a complex fermion in $2d$ has the central charge $c=1$, we conclude that the relevant line bundle on $\M_g$ is also characterized by
\eqn\fermion{
c=-\sigma(M^{(4)})/8.
}

\subsec{Using the anomaly polynomials in  $4d$, $6d$, and $2d$ }

Let us now derive  both results \selfdual, \fermion\ in a unified manner,
in a way applicable to arbitrary $6d$ parent theory.
A line bundle $\L$ over $\M_g$ can be specified by its first Chern class $c_1(\L)$.
Therefore, we would like to compute it starting from the anomaly polynomial of the $6d$ theory.

In general, suppose that we have a family of $2k$ dimensional manifolds $X^{(2k)}(\m)$ parametrized by $\m \in \M$.
Then, the partition function $Z(\m)$ on $X^{(2k)}(\m)$ is a section of a line bundle $\L$ on $\M$.
The anomaly polynomial $\A_{2k+2}$ of a $2k$-dimensional theory is defined on the total space $Y$ of the fiber bundle $X^{(2k)} \to Y \to \M$
in which the fiber over $\m \in \M$ is $X^{(2k)}(\m)$,
and the first Chern class of $\L$ is given by
\eqn\fst{
c_1(\L) = \int_{X^{(2k)}(\m)} \A_{2k+2}.
}
This is the defining property of the anomaly polynomial $\A_{2k+2}$.

Now, our strategy is to consider the $6d$ theory on a six-manifold $N^{(6)}= M^{(4)}\times \Sigma^{(2)}$ and consider the reduction either on $M^{(4)}$ to $2d$, or on $\Sigma^{(2)}$ to $4d$ .  This way we use the anomaly polynomial $\A_8$ of the parent $6d$ theory and compute $c_1(\L) $ using two different orders and compare them.
Specifically, compactifying first on a four-manifold $M^{(4)}$ and then on a two-manifold  $\Sigma^{(2)}$ we find
\eqn\firstcompact{
\A_8 \to \A_4 \to c_1(\L)~.
}
Alternatively, compactifying first on a two-manifold $\Sigma^{(2)}$ and then on a four-manifold $M^{(4)}$ we find
\eqn\secondcompact{
\A_8 \to \A_6 \to c_1(\L)~.
}
They must give the same answer.

First, we consider the process $\A_4 \to c_1(\L)$.
We start by discussing a slightly more explicit expression for the Chern class of the line bundle $\L$ of central charge $c$ over $\M_g$.
We consider the universal bundle $\Sigma^{(2)}\to \U\to \M_g$, whose fiber is the Riemann surface $\Sigma^{(2)}$ parameterized by $\M_g$ itself.
The anomaly polynomial of a $2d$ theory with chiral central charge $c$ is $ - c \, p_1/24$.
As mentioned above, the Chern class of the line bundle $\L$ in which the partition function takes values is then obtained as
\eqn\twod{
c_1(\L)= - {c\over 24}\int_{\Sigma^{(2)}}  p_1(\U)  = - {c\over 24} \int_{\Sigma^{(2)}} c_1(T\Sigma^{(2)})^2 =:  {c\over 2}\lambda
}
where $T\Sigma^{(2)}$ is the tangent bundle of the fiber manifold $\Sigma^{(2)}$,
and $\lambda  =  - {1\over 12} \int_{\Sigma^{(2)}} c_1(T\Sigma^{(2)})^2$ is the Chern class of the Hodge bundle.

Next consider the process $\A_6 \to c_1(\L)$.
Suppose that there is a term in the anomaly polynomial of the $4d$ theory of the form
\eqn\mixed{
\A_6 \supset  \omega \wedge  {p_1(M^{(4)}) \over 3}
}
where $\omega$ is a two-form on $\M_g$ and $M^{(4)}$ is the spacetime of the $4d$ theory.
Then we have
\eqn\coneL{
c_1(\L)=\omega \int_{M^{(4)}}  {p_1(M^{(4)}) \over 3} ={ \sigma} \omega ~.
}
Comparing the above two results, we get
\eqn\comparetwo{
{\sigma } \omega =  {c\over 2}\lambda ,
}
which explains why $c$ is proportional to $\sigma$ in general.

Now we compute $\A_8 \to \A_6$.  Recently, the paper \TachikawaAUX\ discussed the
mixed term in the anomaly polynomial \mixed\ and showed how to compute it.
The point was as follows. We want to compute the lower-dimensional anomaly $\A_6$ by integrating the higher-dimensional anomaly $\A_8$
over the internal Riemann surface as
\eqn\Asix{
\A_6 = \int_{\Sigma^{(2)}} \A_8~.}
In this computation, we need to regard $\Sigma^{(2)}$ as forming the universal bundle $\U$ over $\M_g$.
This then generates mixed terms of the form \mixed~where $\omega$ is proportional to  $\lambda $.

One finds by a short computation that
\eqn\poneptwo{
-{1 \over 12} \int_{\Sigma^{(2)}} p_1(N^{(6)})^2 = 2 \lambda  p_1(M^{(4)})~ ,\qquad
-{1 \over 12} \int_{\Sigma^{(2)}} p_2(N^{(6)}) =  \lambda  p_1(M^{(4)})~,}
where $N^{(6)}$ is the spacetime of the $6d$ theory.
Therefore, the $6d$ anomaly polynomial
\eqn\Aeight{
\A_8= - {a p_1^2 + b p_2 \over 12 \cdot 3}
}
leads to the $4d$ anomaly
\eqn\sixDanom{
\A_6= (2a+b) \lambda  \wedge {p_1(M^{(4)}) \over 3}~.
}
Combining the above results,
the $4d$ theory obtained from the $6d$ theory compactified on $\Sigma^{(2)}$ has an anomaly of the duality group specified by the central charge
\eqn\result{
c= 2(2a+b) {\sigma }.
}

The same computation can be phrased in a slightly different, but essentially the same way by considering the process $\A_8 \to \A_4$:
\eqn\Afoureight{
\A_4=\int_{M^{(4)}} \A_8~.}
One finds
\eqn\poneptwoN{\eqalign{
&{1 \over 3} \int_{M^{(4)}} p_1(N^{(6)})^2=  2  p_1(\U) \int_{M^{(4)}} {p_1 \over 3}  = 2 {\sigma } p_1(\U)~,\cr
&{1 \over 3}  \int_{M^{(4)}} p_2(N^{(6)}) = p_1(\U) \int_{M^{(4)}} {p_1 \over 3}= {\sigma } p_1(\U)~.}}
We now use \twod~to convert them to the effective $2d$ central charge.
We obtain the same result \result.

We apply the formula \result\ to the known anomaly polynomial of a self-dual field and that of a chiral fermion.
Neglecting the Green-Schwarz contribution to the anomaly polynomial from the self-dual field,\foot{
Suppose that the 3-form field strength $H$
of the self-dual field (satisfying $\star H = H$) has the equation of motion $dH/ 2\pi = \alpha p_1(N^{(6)}) $
for some constant $\alpha$. Then it contributes to the anomaly polynomial $\A_8$ as ${\alpha^2 \over 2}p_1(N^{(6)})^2 $.
This contribution is matched by the Green-Schwarz contribution of the compact scalar in $4d$
obtained by integrating the two-form field $B$ over the Riemann surface.
}
we have
\eqn\Afermionsd{
\A_8^{\rm fermion}={7 p_1^2-4p_2\over 2^7 \cdot 3^2 \cdot 5}~,\qquad
\A_8^{\rm s.d.tensor}={p_1^2-7p_2\over 2^3 \cdot 3^2 \cdot 5}}
from which we easily re-obtain
\eqn\ccompare{
c({\rm 4d\ fermion\ coupled\ to}\ \M_g)=-{\sigma\over 8}~,\qquad
c({\rm Maxwell})={\sigma}~.}
This is the same result as in \selfdual, \fermion.

Before proceeding, we mention that the analysis presented in this section can be readily extended to the study of the anomaly of the duality group of $(2n-2)$-dimensional theory obtained by compactifying a $2n$-dimensional theory on 2-dimensional surfaces $\Sigma^{(2)}$ for general $n$.
This would allow us, for example, to reproduce the anomalous phase under $SL(2,\Z)$ of the $2d$ theories obtained by compactifying $4d$ ${\cal N}{=}1$ theories on $T^2$, observed in \ClossetBSE.
It would be interesting to work out the details.

\subsec{Extension of the space of couplings by the gravitational theta angle}
Let us briefly mention how we can make the partition function a function on $\N_g=(\S^1 \times \T_g)/\Gamma_g$
by introducing the gravitational theta angle $\theta_\grav$.

The line bundle $\L$ on $\M_g$ and hence the anomaly can be described by some $\phi_\tau(g)$ for $\tau \in \T_g$ and $g \in \Gamma_g$,
as explained in the general discussion in the previous section. Moreover, we saw that the anomaly is proportional to $\sigma$,
so this $\phi_\tau(g)$ should be represented as $\phi_\tau(g) = (\psi_\tau(g))^{\sigma / 16}$ for some $\psi_\tau(g)$,
which is independent of the four-manifold. Then we define the transformation of $\theta_\grav$ as
\eqn\generaltransf{
g \cdot (\tau, \theta_\grav) = (g \cdot \tau,\  \theta_\grav + i \log \psi_\tau(g) ).
}
As in the previous section, one can check that the partition function is invariant under $\G = \Gamma_g$
and can be regarded as a function on the extended space $\N_g=(\S^1 \times \T_g)/\Gamma_g$.
This is again an $\S^1$ bundle over $\M_g=\T_g/\Gamma_g$.

A difference from the genus one case is that $\psi_\tau(g)$ must depend on $\tau$ for a large enough genus.
The reason is as follows. Suppose otherwise. Then the line bundle $\L$ would have a flat connection
by taking the covariant derivative on $\L$ to be just the ordinary derivative $\partial / \partial \tau$.
Then the first Chern class $c_1(\L)$ would be a torsion and in particular it would be zero at the level of de~Rham cohomology.
However, this contradicts the fact that $c_1(\L)$ can be described as the image of the map
$H^2(B\Gamma_g, \Z)  \to H^2(B\Gamma_g, \R)$ for $g \gg 1$. Therefore, there is no such flat connection on $\L$
and $\psi_\tau(g)$ must depend on $\tau$ for $g \gg 1$.

\newsec{Superconformal theories}

Finally let us make a brief comment on the $4d$ \Nequals2 superconformal theories.
A parallel discussion can be given for $2d$ \Nequals{(2,2)} theories, so we will not repeat it.
In this section we follow the notation used in \GomisYAA.

As above, $\M$ is the space of exactly marginal couplings.
The Zamolodchikov metric on $\M$ is known to be K\"ahler.
The relevant background fields are the fields of the
\Nequals2 supergravity multiplet and their counterterm is
\eqn\Ntwosu{
t {1\over 192\pi^2}  \int d^4xd^4\theta {\cal E}(\Xi-W^{\alpha\beta}W_{\alpha\beta}) + c.c.~.
}
The real part and the imaginary part of $t$ multiply the Euler density and the Pontryagin density, respectively,
and hence it contributes to the action as ${\rm Re}(t) \chi$ and ${\rm Im}(t) \sigma$ with some numerical coefficients.
In particular, ${\rm Im}(t)$ is proportional to the gravitational theta angle $\theta_\grav$.
The term \Ntwosu\ also contains the background theta-angle for $SU(2)_R$.
There is another term in supergravity which is proportional to $ \int d^4xd^4\theta {\cal E} W^{\alpha\beta}W_{\alpha\beta}$,
but that term does not change the qualitative discussion below.

Due to the existence of $t$, we have a fibration
\eqn\Ntwofib{
\C/\Z \to \N \to \M ~,}
where $\C/\Z$ is the space parameterizing $t$.

The K\"ahler transformation of the K\"ahler potential $K_\M$ of the conformal manifold $\M$ are given by
\eqn\Kahlertr{
K_\M \to K_\M +F+\bar F}
and it is accompanied by the shift  \GomisYAA
\eqn\Kahlerts{
t\to t + {F\over 2}~.}
This means that the first Chern class of the fibration $\N\to \M$ is given by the K\"ahler class of $\M$.

For class S theories, obtained by putting a $6d$ \Nequals{(2,0)} theory on a Riemann surface of genus $g$,
the conformal manifold $\M$ is the moduli space $\M_g$ of genus-$g$ Riemann surfaces \GaiottoWE.
We can compute the Chern class of the fibration from the $6d$ anomaly polynomial as discussed in \TachikawaAUX\
and also in a previous section in this note.
This can then be used to compute the K\"ahler class.\foot{%
Precisely speaking, more information is needed in order to fully specify the theory \refs{\AharonyHDA,\TachikawaHYA}.  Depending on which $\N=(2,0)$ theory we compactify,
one needs to specify a Lagrangian subgroup of the first cohomology group of the Riemann surface on which the $6d$ theory is compactified.
This reduces the duality group from $\Gamma_g$ to a certain subgroup, which preserves this data.
We then have to study the line bundle on the moduli space divided by that subgroup.
This is still specified by the effective $2d$ central charge.
This point is perfectly analogous to the situation we mentioned in a previous footnote when we discussed chiral fermions.
In special cases, like the $\N=(2,0)$ theory of type $U(N)$  (i.e.\ a stack of $N$ M5-branes including the center of mass degrees of freedom) or type $E_8$
 the duality group is not reduced.
 This is associated with the fact that the lattice of charges in these theories is self-dual.
}

In \GomisYAA, it was suggested that the K\"ahler potential should be globally well-defined on the space fully parameterizing a family of superconformal theories.
The discussion in this note means that this space, fully parameterizing a family of superconformal theories, should be taken as $\N$ including the background coupling $t$, instead of $\M$, which includes only the dynamical couplings.\foot{%
It was shown in \DonagiVWH\ that for large enough $g$ the K\"ahler class of $\M_g$ is nontrivial,
even when restricted to the points where there is no enhancement of symmetry of the curve $\Sigma^{(2)}_g$.
Therefore, the K\"ahler potential is not globally well defined on $\M_g$.
Also, the authors of \DonagiMHD\ studied the complex structure moduli spaces of a few compact Calabi-Yau manifolds,
which are the B-model conformal manifolds of the corresponding $2d$ \Nequals{(2,2)} supersymmetric sigma models.
It was found there that the K\"ahler classes of these spaces are torsion but can be nontrivial.
}
And indeed, on $\N$, there is a natural globally well-defined ``K\"ahler potential''
\eqn\KahleronN{
K_\N := -2t -2\bar t + K_\M~.}
We put the quotation marks around the term ``K\"ahler potential'', since $g_{t\bar t}=\partial_t \bar\partial_{\bar t}K_\N=0$ is degenerate.
This is physically expected, since the direction along the background coupling $t$ has zero Zamolodchikov metric,
because in \Ntwosu\ $t$ is the coefficient of a term constructed purely out of background supergravity fields.

Finally, as we said in the introduction, instead of adding $t$ as a coordinate and view the theory as a function on $\N$, we also have the option to keep $t$ fixed and mod out the parameter space
only by the subgroup of the duality group that preserves $t$.
Then the theory and the partition function are single-valued on a multiple cover $\widehat\M$ of $\M$.

The discussion in this section is similar to the analysis of ``field-dependent Fayet-Iliopoulos terms'' \DineXK\ and possible compact cycles in the moduli space of string theory.  As discussed in \refs{\KomargodskiPC,\KomargodskiRB,\SeibergQD}, ordinary FI-terms and non-trivial K\"ahler class in the moduli space are inconsistent in a theory of gravity (unless they are properly quantized).  Apparent FI-terms and apparent non-trivial K\"ahler classes are accommodated in string theory by including the coupling to additional moduli $t$ (e.g.\ the dilaton) of the form \KahleronN.  With this additional field there are no non-trivial classes.

\bigskip

\goodbreak
\noindent {\bf Acknowledgments:}

The authors would like to thank C.~C\'ordova, R.~Donagi, and D.~Morrison for numerous discussions on these topics and in particular about the existence of closed two-cycles in $\M$ and about the stimulating paper \DonagiVWH.
Useful conversations with C.-T.~Hsieh, Z.~Komargodski, R.~Thorngren and E.~Witten are also acknowledged.
The work of NS was supported in part by DOE grant DE-SC0009988.
The work of YT was partially supported  by JSPS KAKENHI Grant-in-Aid (Wakate-A), No.17H04837
and JSPS KAKENHI Grant-in-Aid (Kiban-S), No.16H06335,
and also by WPI Initiative, MEXT, Japan at IPMU, the University of Tokyo.
The work of KY is supported in part by the WPI Research Center Initiative (MEXT, Japan),
and also supported by JSPS KAKENHI Grant-in-Aid (Wakate-B), No.17K14265.

Any opinions, findings, and conclusions or recommendations expressed in this material are those of the authors and do not necessarily reflect the views of the funding agencies.

\appendix{A}{Symmetry protected topological phases of duality groups}

It is widely believed that we can describe an anomaly by a symmetry protected topological (SPT) phase in one-higher dimension.
We would like to test this idea in the case of the duality groups $\Gamma_g$.

Let $B\Gamma_g$ be the classifying space of the duality group $\Gamma_g$.
In $d$-spacetime dimensions, a global anomaly may be classified by the torsion part
of the bordism group $\Omega^{\rm spin}_{d+1}(B\Gamma_g   )_{\rm tor}$, while a perturbative anomaly (in the sense that it is not a torsion)
may be classified by the non-torsion part of the bordism group in $(d+2)$-dimensions
$\Omega^{\rm spin}_{d+2}(B\Gamma_g  ) \otimes \R$ \refs{\KapustinDXA,\FreedRQQ,\YonekuraUFJ}.
In this paper we only consider spin manifolds and restrict our attention to the spin bordism group.

\subsec{Genus one case}

The anomaly discussed in the main text
is a mixed anomaly between the signature and $SL(2,\Z)$,
and the signature density has dimension 4.
Let $a \in H^1(BSL(2,\Z), U(1)) =\Z_{12}$ be the generator of $\Z_{12}$.
From the anomaly shown in \newtrans,
we have seen that $T$ gives the phase factor $e^{-\pi i \sigma /12} $,
while $S$ does not have a phase.
Hence, $T$ has an anomaly $e^{-\pi i \sigma /12}$ and
a reasonable guess is that the five-dimensional SPT phase characterizing the anomaly
is schematically given by $-\pi i \int  (a \wedge p_1 /3 )/12 $ where $p_1/3$ is the cohomology element corresponding to the signature $\sigma$.

We will show below that
\eqn\total{
0\to \Z_6\to \Hom(\Omega_5^{\rm spin}(BSL(2,\Z),U(1))=\Z_{36}\to \Z_6\to 0
}
where the $\Z_6$ quotient  on the right hand side classifies the mixed anomaly between $SL(2,\Z)$ and the signature,
and the $\Z_6$ subgroup on the left hand side classifies the pure $SL(2,\Z)$ anomaly.

This means that there is a $5d$ SPT phase $A$
characterizing the $4d$ mixed $SL(2,\Z)$-gravity anomaly
which satisfies  $A^{\otimes 6}=1$ modulo pure $SL(2,\Z)$ anomaly.
Let us explicitly construct a $4d$ theory having this anomaly.
We take a $6d$ Weyl fermion coupled to a $U(1)$ background field and
we compactify it on a $2d$ torus with the $U(1)$ bundle identified with the spin bundle on the torus. Then,
the total system does not depend on the spin structure of the $2d$ torus and the group acting on this torus is $SL(2,\Z)$ (i.e. not its spin cover).\foot{
This is a ``topological twist", although we are discussing a non-supersymmetric theory.}
We get a $4d$ fermion system coupled to $\tau \in \H$. If we first compactify the $6d$ fermion on a four-manifold as in section~4,
we get the net $-\sigma/8$ copies of the $b,c$ ghost system with weight $(h_b,h_c)=(0,1)$, and the chiral central charge is $c = (-2)(-\sigma/8)=\sigma/4$.
The corresponding line bundle on $\M=\H/SL(2,\Z)$ is $\L=L^{\otimes c/2}=(L^{\otimes 2})^{\otimes (\sigma/16)}$ where $L$ is the generator of the line bundles on $\H/SL(2,\Z)$,
$L^{\otimes 12}=1$.
This $\L$ has order 6, $\L^{\otimes 6}=1$.

Before proceeding, let us discuss the Maxwell theory, which had $c=\sigma$ as we saw.
This means that its $SL(2,\Z)$ anomaly corresponds to  $A^{\otimes 4}$.
This implies that the Maxwell theory has an anomaly $a \in \Z_{36}$ such that $a \equiv 4\ \hbox{mod}\ 6$,
when we include the pure $SL(2,\Z)$ anomaly.
In particular, we see  that three copies of the Maxwell theory necessarily has a nontrivial pure $SL(2,\Z)$ anomaly corresponding to $12\in \Z_{36}$ or $30 \in \Z_{36}$.
It would be interesting to determine exactly which anomaly the Maxwell theory has.

\subsec{Higher genus case}
Before proving \total, let us discuss the case of higher genus Riemann surfaces.
One starts from an element $\omega$ in $H^2(B\Gamma_g,\Z)$.
This element $\omega$ can be regarded as the first Chern class of a line bundle on $B \Gamma_g$, and
let $a$ be a local connection of the line bundle such that $\omega=d a$ at the level of de~Rham cohomology.
We consider the five-dimensional theory which is proportional to
\eqn\mixedCS{
\int a \ p_1.
}
This is the five dimensional SPT phase of our interest.
It has the anomaly polynomial of degree-6 proportional to
\eqn\anomalypol{
\omega\ p_1.
}
This gives an element of $\Omega^{\rm spin}_{6}(B\Gamma_g) \otimes \R$ if $\omega$ is not a torsion.

Note that, up to orbifold singularities that only affect torsions,
the moduli space of Riemann surfaces $\M_g$ may be regarded as a model for a classifying space $B\Gamma_g$.
Therefore, $\omega$ may be seen as a 2-form on $H^2(\M_g,\Z)$.
Therefore, two-forms on the moduli space $\M_g$
are directly related to the mixed anomaly between the duality group $\Gamma_g$ and the gravity.
This is exactly what we found in section 4,
and was the new type of anomaly discussed in \TachikawaAUX.

\subsec{Computation of $\Omega^{\rm spin}_5(BSL(2,\Z))$}

Let us first recall how $H_*(BSL(2,\Z),\Z)$ is computed (see e.g.~Chap.~II.7 of \BrownCohomology).
We first note that $SL(2,\Z)$ is an amalgam of $\Z_4$ and $\Z_6$ over its common subgroup $\Z_2$.
This means the following:
the group $SL(2,\Z)$ is obtained by gathering generators and relations of
\eqn\ZfourZsix{
\Z_4=\vev{S \mid  C:=S^2, C^2=1}, \quad
\Z_6=\vev{T^{-1}S \mid C:=(T^{-1}S)^3, C^2=1}
}
by identifying the common subgroup $\Z_2=\vev{C \mid C^2=1}$.
Correspondingly, the classifying spaces can be chosen so that
\eqn\class{
BSL(2,\Z)=B\Z_4 \cup B\Z_6,\qquad
B\Z_4 \cap B\Z_6 = B\Z_2.
}
Then we have the Mayer-Vietoris long exact sequence
\eqn\MVhomology{
H_n(B\Z_2,\Z) \to
H_n(B\Z_4,\Z)\oplus H_n(B\Z_6,\Z) \to
H_n(BSL(2,\Z),\Z) \to
H_{n-1}(B\Z_2,\Z).
}
We now use the standard fact that $H_n(B\Z_k,\Z)$ for $n>0$ is given by $\Z_k$ for odd $n$ and by $0$ for even $n$ to conclude that $H_n(BSL(2,\Z))$  for $n>0$ is given by $\Z_{12}$ for odd $n$ and by $0$ for even $n$.

We now move on to the computation of $\Omega^{\rm spin}_5(BSL(2,\Z))$.
Since the spin bordism is a generalized homology theory, we similarly have the long exact sequence
\eqn\MVspinbordism{
\Omega^{\rm spin}_n(B\Z_2) \to
\Omega^{\rm spin}_n(B\Z_4)\oplus \Omega^{\rm spin}_n(B\Z_6) \to
\Omega^{\rm spin}_n(BSL(2,\Z)) \to
\Omega^{\rm spin}_{n-1}(B\Z_2).
}
We use the case $n=5$.
By Smith isomorphism (see e.g.~Lemma 6 of \KirbyTaylorPaper\ or the section 6 of \KapustinDXA\ or more explicitly Table~1 of the latter paper), we know
$\Omega^{\rm spin}_n(B\Z_2)\simeq\Omega^{{\rm pin}-}_{n-1}({\rm pt} )\oplus \Omega^{\rm spin}_n({\rm pt})$, and therefore $\Omega^{\rm spin}_5(B\Z_2)=0$ and $\Omega^{\rm spin}_4(B\Z_2)=\Z$.
As $\Omega^{\rm spin}_5(BG)$ for $G=\Z_4$, $\Z_6$, and $SL(2,\Z)$ are clearly torsion from the Atiyah-Hirzebruch spectral sequence (AHSS), we have
\eqn\rrr{
\Omega^{\rm spin}_5(B\Z_4)\oplus \Omega^{\rm spin}_5(B\Z_6)  \simeq
\Omega^{\rm spin}_5(BSL(2,\Z)).
}
Now $\Omega^{\rm spin}_5(B\Z_4)$ was computed to be $\Z_4$ in \GilkeyBotvinnik,
and $\Omega^{\rm spin}_5(B\Z_3)$ was computed to be $\Z_9$ in \RosenbergII.
From AHSS again, we have $\Omega^{\rm spin}_5(B\Z_6)=\Omega^{\rm spin}_5(B\Z_3) \oplus \Omega^{\rm spin}_5(B\Z_2)=\Z_9$.\foot{%
For a detailed readable account for physicists on $\Omega^{\rm spin}_5(B\Z_n)$ for arbitrary $n$,
see \Hsieh.
}
We therefore conclude
\eqn\zzz{
\Omega^{\rm spin}_5(BSL(2,\Z))=\Z_{36}.
}

Since this contains both the pure $SL(2,\Z)$ part and the mixed $SL(2,\Z)$-gravity part, we would like to separate them apart.\foot{%
Recall that the AHSS is a spectral sequence converging to $\Omega_{p+q}^{\rm spin}(BG)$
whose $E^2$ page is $E^2_{p,q}=H_p(BG,\Omega_q^{\rm spin}({\rm pt}))$.
Roughly speaking, the mixed $G$-gravitational part is $(p,q)=(1,4)$ and the pure $G$ part is $(p,q)=(5,0)$.
But unlike at the level of the anomaly polynomial, the entire $\Omega_5^{\rm spin}(BG)$ is not necessarily the direct sum.
}
We consider two AHSSs at the same time, one for $G=SL(2,\Z)$ and another for $G'=\Z_2$,
and use its compatibility with the homomorphism $G'\to G$ obtained by sending $G'$ to the center of $G$,
and also the homomorphism $G\to G'$ given by the Abelianization $SL(2,\Z)\to \Z_{12}$ composed with the standard $\Z_{12}\to \Z_2$.
The $E^2_{p,5-p}$ for $G=SL(2,\Z)$ and $G'=\Z_2$ are given by
\eqn\ooo{
0,\
\Z_{12},\
0,\
\Z_2,\
\Z_2,\
\Z_{12};
\qquad
0,\
\Z_{2},\
0,\
\Z_2,\
\Z_2,\
\Z_{2},
}
respectively for $p=0,1,2,3,4,5.$
We note that $E^\infty_{p,5-p}$ for $G'=\Z_2$ is empty, since $\Omega^{\rm spin}_5(B\Z_2)=0$.
This fact tells us how various differentials act for $G'=\Z_2$.
This knowledge can then be transferred to the case $G=SL(2,\Z)$ by the homomorphisms $G\to G'$ and $G'\to G$  discussed above.
The result is that $E^\infty_{p,5-p}$ for $G=SL(2,\Z)$ are given by
\eqn\iii{
0,\ \Z_6,\ 0,\ 0,\ 0,\ \Z_6
} for $p=0,1,2,3,4,5$.
Comparing with \zzz, we conclude that we have the sequence
\eqn\homototal{
0\to \Z_6\to \Omega_5^{\rm spin}(BSL(2,\Z))=\Z_{36}\to \Z_6\to 0
}
where
the $\Z_6$ subgroup on the left hand side is $E^\infty_{1,4}=\Z_6$ for the mixed anomaly,
while
the $\Z_6$ quotient  on the right hand side is $E^\infty_{5,0}=\Z_6$  for the pure anomaly.
By taking the Pontryagin dual of the entire sequence,
we finally obtain the statement \total\ we used in Appendix A.1,
where the $\Z_6$ quotient  on the right hand side is for the mixed anomaly
while the $\Z_6$ subgroup on the left hand side is for the pure anomaly.

\appendix{B}{Theta angle, its periodicity in $4d$ $SU(N)$ gauge theory, and axions}

The purpose of this appendix is to demonstrate some of the phenomena we discussed in the main part, using a pure gauge $4d$ $SU(N)$  theory.

We start by reviewing the analysis of this system and its $\theta$ dependence from \GaiottoYUP.
Traditionally the system depends on a $2\pi$-periodic $\theta$-parameter.
So using the notation of this paper we can say that $\M=\S^1$ is parameterized by \eqn\Mgauge{\M: \qquad \theta \sim \theta+2\pi~.}
This system has a $\Z_N$ one-form global symmetry \refs{\KapustinGUA,\GaiottoKFA} and we can couple it to a background two-form gauge field $B$.
As we will now review, the periodicity of $\theta$ is different when such nonzero $B$ is included.  For simplicity, we will focus on spin manifolds, where it is $\theta \sim \theta + 2\pi N$. (It is straightforward to extend our discussion to nonspin manifold where for $ N$ even $\theta \sim \theta + 4\pi N $ and for $ N$ odd  $\theta \sim \theta + 2\pi N $.)
In the notation of this paper $\N=\S^1$, but its size is larger than that of $\M=\S^1$.
After reviewing this result we will examine its consequences when $\theta$ is a dynamical field, an axion.

The $SU(N)$ gauge theory has nontrivial observables associated with the partition function with twisted 't Hooft boundary conditions.  These are obtained by viewing the theory as a $PSU(N)$ gauge theory, whose bundles $P$ are characterized by the second Stiefel-Whitney class $w_2(P)$.  Then, the partition function with fixed $w_2(P)$ are nontrivial observables of the $SU(N)$ theory.  One way to think about these observables is to note that the $SU(N)$ gauge theory has a $\Z_N$ one-form global symmetry, which we can couple to a classical, background, two-form gauge field $B^{\rm discr}$.  This has the effect of setting $w_2(P)=B^{\rm discr}$.

As in \KapustinGUA, we describe the background gauge field for the $ \Z_ N$ one-form global symmetry $B^{\rm discr}$ using a continuum notation.
We embed the original $SU(N)$ theory in a $U(N)$ gauge theory with field strength $f'$ and add a one-form gauge symmetry to eliminate the added $U(1)$ field.
Specifically, we express the Lagrangian for the $SU(N)$ field $f$ in terms of $f'$ and add a Lagrange multiplier $U(1)$ two-form gauge field $u$ with the coupling
\eqn\ufpL{{1\over 2\pi} u (dC-\Tr f')~,}
which sets the value of the $U(1)\subset U(N)$ gauge field  $\Tr\ a'=C $ (up to a gauge transformation) in terms of a classical $U(1)$ gauge field $C$.
We also define the background gauge field $B $ in terms of
\eqn\BCd{NB \equiv dC~ .}
The one-form gauge symmetry with gauge parameter $\lambda^{(1) }$ acts as\foot{Here and below the superscript in parenthesis denotes the degree of the form.}
\eqn\oneformg{\eqalign{
&a'\to a'+\lambda^{(1) }\I~,\cr
&C\to C+N\lambda^{(1) }~,\cr
&B\to B+d \lambda^{(1) }~,}}
where $\I$ is an $N\times N$ unit matrix.

The gauge invariant information contained in $B$ is as follows.
By integrating ${N \over 2\pi} B = {1 \over 2\pi} dC$ over two cycles
we get integer values ${N \over 2\pi}\oint B \in \Z $.
However, the above gauge transformation by $\lambda^{(1) }$
(which has the flux quantization condition ${1\over 2\pi} \oint d \lambda^{(1) } \in \Z $) can shift those integers by multiples
of $N$.
Therefore, the gauge invariant information is those integers mod $N$, i.e.\ ${N \over 2\pi} \oint B \in \Z_N $, and effectively, $B$ is a two-form $\Z_N$ gauge field.
It can be written as a discrete integer-valued field $B^{\rm discr}={N \over 2\pi} B$.

We can also add to the Lagrangian the counterterm in the background field
\eqn\invcop{ {p \over 4\pi N} \int dCdC = {p N\over 4\pi } \int BB~,}
which is invariant under the transformation by $\lambda^{(1) }$ if $p \in \Z$.
The value of $p$ is meaningful only mod $N$,
$p\sim p+N$, because $ {1 \over 2(2\pi)^2} \int dCdC  \in \Z$ on a spin manifold.

In terms of the discrete $\Z_N$ background field $B^{\rm discr}={N \over 2\pi} B$, \invcop\ is
${p\pi \over N} \int B^{\rm discr}\cup B^{\rm discr} $.  (At the level of cohomology with $\Z_N$ coefficients, ${1 \over 2} B^{\rm discr}\cup B^{\rm discr}$
should be replaced by Pontryagin square.)
In terms of $PSU(N)$ gauge fields with a bundle $P$ the coupling \invcop\ is the counterterm ${p\pi \over N} \int w_2(P)\cup w_2(P) $ (again, with the Pontryagin square.)

Then the $\theta$-term in the $SU(N)$ Lagrangian becomes
\eqn\thetater{{\theta \over 8\pi^2} \int \left(\Tr (f'f') - {1\over N }dCdC\right)~.}
Here the $C$ dependent term is set such that it removes the contribution of $\Tr\ a'$ in the first term.
Equivalently, it makes \thetater\ gauge invariant under the one-form symmetry \oneformg.
However, this term is not $2\pi$-periodic in $\theta$.
As a result, a shift of $\theta$ by $2\pi$ does not leave the action invariant, but shifts it by
$-{1\over 4\pi N} \int dCdC = -{N\over 4\pi} \int BB $.
This means that it shifts $p\to p-1$.
(Recall that we limit ourselves to spin manifolds.  For the more general case see \refs{\KapustinGUA,\GaiottoKFA,\GaiottoYUP}.)
Therefore, we cannot identify the theory with $\theta$ with the theory with $\theta +2\pi$; they have different counterterms.
More explicitly, the identification is
\eqn\idneti{(\theta,p) \sim (\theta + 2\pi, p+1)~.}
Correspondingly, we identify $\N$ as
\eqn\Ngauge{\N: \qquad
\theta \sim \theta+2\pi N   ~.}
This shows that although both $\M$ and $\N$ are circles, $\N$ is a larger circle.
Namely, it has the structure of a fiber bundle ${\cal F} \to \N \to \M$, and the fiber ${\cal F}$ is $ \Z_{N}$
which is parameterized by the integer $p \in {\cal F} = \Z_N$.

In terms of constrained $PSU(N)$ fields, nontrivial $w_2(P)$ makes the instanton number fractional and hence the larger period of $\theta$.

The distinction between $\M$ and $\N$ leads to a natural question about the system with $\theta$ a dynamical field, an axion.
We let $\theta$ be spacetime dependent, add a kinetic term, e.g.\ $\half f_\theta ^2 (\partial\theta)^2$, and integrate over it.
Then, what are the allowed periodicities of $\theta$?  Are they multiples of $2\pi$ as in $\M$ \Mgauge, or are they multiples of $2\pi N$ as in $\N$ \Ngauge?

First, let us set the background fields $NB=dC$ to zero.
Now we can let the field $\theta$ be $2\pi$-periodic.
One way to think about it is by starting with a noncompact field $\theta$ and noticing that the system has a {\it global} symmetry $\Z$, whose generator shifts $\theta$ by $2\pi$.
In the notation of the main text (see section 2 and 3),
this means that we consider the space $\widehat{\M} = \R$ as a space in which $\theta$ takes values, and
consider the group $\G = \Z$ that acts on $\widehat{\M}$.  However, unlike the discussion there, here $\theta$ is a dynamical field rather than a background field and therefore $\G = \Z$ is a global symmetry.

Then, when we restore the background fields we realize that only shifts by multiples of $2\pi N$, as in \Ngauge, do not affect the background fields.  And shifts by other multiples of $2\pi$ do affect them.
We can interpret this to mean that there is a mixed 't Hooft anomaly between the global symmetry group $\G=\Z$ shifting $\theta$ and the one-form $\Z_N$ global symmetry.
This is an example of the general discussion in section~3.1 when $\theta$ is a field.

There is clearly no problem gauging a subgroup of the $\Z$ shift symmetry and identify $\theta \sim \theta +2\pi N$, as in \Ngauge.
But can we identify it further and make $\theta \sim \theta+2\pi$?  What is the effect of the 't Hooft anomaly on making such an identification?

The key point in the answer to this question is the fact that the system with a dynamical compact $\theta$ has a $U(1)$ two-form global symmetry \GaiottoKFA.
Its generator is the line operator $U_\alpha = e^{i\alpha \oint d\theta}$ and its charged objects are domain walls between two values of $\theta$ that differ by a multiple of $2\pi$.
We can couple this symmetry to a background three-form gauge field $A^{(3)}$ by adding to the Lagrangian ${1\over 2\pi} \theta dA^{(3)}$.  This gauge field has gauge freedom $A^{(3)} \to A^{(3)} + d\lambda^{(2)}$. (Soon we will modify this gauge transformation and this integer will not be gauge invariant.)
Then, we can replace the $\theta$ term \thetater\ by
\eqn\thetaterm{{1\over 8\pi^2} \int \theta\left( \Tr (f'f')  + 4\pi dA^{(3)}\right)~.}
It is invariant under $\theta \to \theta +2\pi$.
Because of the presence of $A^{(3)}$, it is gauge invariant under the $\lambda^{(1)}$ gauge transformation \oneformg\ provided $A^{(3)}$ transforms as
\eqn\Athreetran{A^{(3)} \to A^{(3)} + d\lambda^{(2)} - {1\over 2\pi} \lambda^{(1)} dC -{N \over 4\pi} \lambda^{(1)}d\lambda^{(1)}~.}
(We should also shift $u$ in \ufpL\ appropriately.)
This means that the gauge invariant field strength of the classical fields is
\eqn\Ffour{F^{(4)}=dA^{(3) }+{1\over 4\pi N} dCdC=dA^{(3) }+{N\over 4\pi } BB~.}

The mixed transformation law \Athreetran\ are characteristic of a higher group structure and the discussion here fits nicely in the framework of \CordovaCVG\ (see also references therein).

Another way to write \thetaterm\ is
\eqn\thetatermm{{1\over 8\pi^2} \int \theta\left( \Tr (f'f') -{1\over N}dCdC \right)  + {1\over 2\pi} \int \theta F^{(4)}~,}
where the first term is the original gauge invariant term \thetater\ and the second term depends on the field strength \Ffour.  Neither term is $2\pi$-periodic in $\theta$, but their sum is.

It should also be pointed out that we can add to the theory another gauge invariant counterterm in the background fields
\eqn\rhoterm{{\rho\over 2\pi}  F^{(4)} }
with constant $\rho$.  The original lack of $2\pi$-periodicity in $\theta$ is replaced by lack of $2\pi$ periodicity in $\rho$.  Now $(\rho,p) \sim (\rho+2\pi, p-1)$.

\listrefs
\bye